# Franckeite: a naturally occurring van der Waals heterostructure

Aday J. Molina-Mendoza,[1,†] Emerson Giovanelli,[2,†] Wendel S. Paz,[1] Miguel Angel Niño,[2] Joshua O. Island,[3] Charalambos Evangeli,[1,§] Lucía Aballe,[4] Michael Foerster,[4] Herre S. J. van der Zant,[3] Gabino Rubio-Bollinger,[1,5] Nicolás Agraït,[1,2,5] J. J. Palacios,[1,*] Emilio M. Pérez,[2,*] and Andres Castellanos-Gomez.[2,*]

[1]Departamento de Física de la Materia Condensada, Universidad Autónoma de Madrid, Campus de Cantoblanco, E-28049, Madrid, Spain.

[2]Instituto Madrileño de Estudios Avanzados en Nanociencia (IMDEA-Nanociencia), Campus de Cantoblanco, E-28049 Madrid, Spain.

[3]Kavli Institute of Nanoscience, Delft University of Technology, Lorentzweg 1, 2628 CJ Delft, The Netherlands.

[4]ALBA Synchrotron Light Facility, Carrer de la Llum 2-26, Cerdanyola del Vallés, Barcelona 08290, Spain.

[5]Condensed Matter Physics Center (IFIMAC), Universidad Autónoma de Madrid, E-28049 Madrid, Spain.

[†]*These authors contributed equally.*

[§]*Present address:* Department of Physics, Lancaster University, Lancaster LA1 4YB, United Kingdom.

*email: juanjose.palacios@uam.es; emilio.perez@imdea.org;

andres.castellanos@imdea.org.

**The fabrication of van der Waals heterostructures, artificial materials assembled by individually stacking atomically thin (2D) materials, is one of the most promising directions in 2D materials research. Until now, the most widespread approach to stack 2D layers relies on deterministic placement methods which are cumbersome when fabricating multilayered stacks. Moreover, they tend to suffer from poor control over the lattice orientations and the presence of unwanted adsorbates between the stacked layers. Here, we present a different approach to fabricate ultrathin heterostructures by exfoliation of bulk franckeite which is a naturally occurring and air stable van der Waals heterostructure (composed of alternating $SnS_2$-like and PbS-like layers stacked on top of each other). Presenting both an attractive narrow bandgap (<0.7 eV) and p-type doping, we find that the material can be exfoliated both mechanically and chemically down to few-layer thicknesses. We present extensive theoretical and experimental characterizations of the material's electronic properties and crystal structure, and explore applications for near-infrared photodetectors (exploiting its narrow bandgap) and for p-n junctions based on the stacking of $MoS_2$ (n-doped) and franckeite (p-doped).**

The demonstration of the deterministic placement of 2D crystals has opened the door to fabricate more complex devices,[1] but more importantly, it has started the investigation to tailoring the properties of designer materials by stacking different 2D crystals to form so-called van der Waals heterostructures.[2] One approach to produce such heterostructures is to use epitaxially grown materials assembled sheet by sheet.[3] This method, however, remains challenging and has only been demonstrated for a few van der Waals heterostructures so far.[4, 5, 6] Another approach is the manual assembly of individual layers obtained by mechanical exfoliation from bulk and the deterministic placement of one layer on top of another.[7, 8, 9, 10, 11] This method also presents issues that remain to be solved, such as controlling the exact crystalline alignment between the stacked lattices and avoiding the presence of interlayer atmospheric adsorbates.

Here, we present the study of ultrathin layers of franckeite, a naturally occurring sulfosalt with a structure formed by alternated stacking of tin disulfide-based ($SnS_2$) and lead sulfide-based (PbS) layers. Interestingly, the individual layers present a larger bandgap than the naturally formed van der Waals heterostructure (<0.7 eV), which is amongst the narrowest found in 2D semiconductors. We also find that franckeite is a p-type material, a very rare feature in two-dimensional semiconductors, only found so far in a few materials such as black phosphorus and tungsten diselenide [12, 13, 14, 15, 16] but, unlike black phosphorus, franckeite is air-stable. We combine density functional theory (DFT) calculations with the experimental characterization of the optical and electrical properties of ultrathin franckeite, which we isolate both by micromechanical cleavage and liquid-phase exfoliation, to offer a complete picture of its unique properties.

**Franckeite crystal and band structure**

Franckeite is a layered material from the sulfosalt family formed by the stacking of pseudohexagonal (H) and pseudotetragonal (Q) layers that interact by van der Waals forces.[17] The Q layer is composed of four atomic layers of sulfide compounds with the formula MX, where M = $Pb^{2+}$, $Sn^{2+}$ or $Sb^{3+}$ and X = S. The H layer consists of octahedrons of disulfide compounds with the formula $MX_2$, where M = $Sn^{4+}$ or $Fe^{2+}$ and X = S. In Fig. 1a we show the crystal structure of the material, indicating the different atomic layers present in the crystal.

In order to shed light on the effect of this alternated stacking of H and Q layers on the electronic properties of franckeite, we perform density functional theory (DFT) calculations of the band structure as implemented with the QuantumEspresso code,[18] based on the crystal structure described above. Apart from the H and Q layers already mentioned, the franckeite structure exhibits a long-range one-dimensional transversal wave-like modulation and a non-commensurate layer match in two dimensions.[17] This will not be taken into account since it would be computationally too expensive and it is expected to only introduce minor corrections to the results.

We first investigate the band structure of individual Q and H layers (Fig. 1b and 1c, respectively). In these calculations we only consider PbS and $SnS_2$ compounds, ignoring the influence of substitutional Sb or Fe atoms (we direct the reader to the Supporting Information for calculations including Sb substitutional atoms). Direct bandgaps of ~1 eV and ~0.7 eV are obtained for the H and Q layers, respectively (Figs. 1b and 1c). The band structure of the bulk crystal formed by alternate stacking of Q and H layers with the same composition as that of the individual layers, results in two sets of bands separated by a small gap of ~0.5 eV (Fig. 1d). Although the Fermi level lies below the valence band, a further investigation of the role played by substitutional Sb atoms (Figs. S1 and S2 in the Supporting Information) shows that the Fermi level is shifted upwards when the material is doped with Sb. These results can be compared with the electronic bandgap measured by means of scanning tunneling spectroscopy (STS) on bulk franckeite crystals, where we observe an electronic bandgap of ~0.7 eV with the valence band edge closer to the Fermi level, further confirming the p-type doping of the material (also confirmed by the thermopower measurements). We direct the reader to the Supporting Information for more details on the STS and thermopower measurements.

In the band structure shown in Fig. 1d, we note that the valence band wavefunctions correspond to the H layer while the conduction band wavefunctions belong to the Q layer, much in analogy with an artificial type-II semiconducting heterostructure.[19] In Fig. 1e we show a projection of the crystal structure with the corresponding Bloch states, where the blue areas represent the Bloch states corresponding to the blue band in Fig. 1d and the red areas represent the Bloch states in the in the red band in Fig. 1d.

**Transmission electron microscopy and micro-XPS of mechanically exfoliated flakes**

Transmission electron microscopy (TEM) of mechanically exfoliated franckeite flakes reveal both the high degree of orientation in the stacking of the H and Q layers and its misfit structure. Fig. 2a shows a low magnification TEM image of a flake with regions of different thicknesses, as inferred from the difference in TEM contrast. The TEM image shows regularly spaced fringes which are due to the corrugation of the crystal induced by the interaction between the misfit Q and H layers, as demonstrated by Makovicky *et al.*[17]

with cross sectional TEM in bulk franckeite. Fig. 2b shows a high resolution (HRTEM) image where the atoms of both the Q and H layers can be resolved. The corresponding selected area electron diffraction (SAED) diagram (Fig. 2c) has consequently been indexed considering franckeite as a stacking of the Q and H layers, respectively described as tetragonal and orthohexagonal.[20] We address the reader to Fig. S4 in the Supporting Information for scanning electron microscopy images of mechanically exfoliated flakes.

We have also characterized mechanically exfoliated franckeite flakes using synchrotron micro-X-ray photoemission spectroscopy (XPS) with lateral resolution of ~20 nm in a photoemission electron microscope (PEEM).[21] The flakes were transferred onto a metallic Pt surface to avoid sample charging. XPS-PEEM spectra of the four main components of franckeite Sb 3d, Sn 3d, S 2p and Pb 4f are displayed in Fig. 2d and e. From the core level fits we distinguish two components for each element: for Sn, we assign the two components to $Sn^{2+}$ and $Sn^{4+}$ from the Q and H structural layers of franckeite, respectively. The Sb 3d core level also presents two components: $Sb^{3+}$ appears in the Q layer with an extra Sb in a different environment (see more details in the Supporting Information, as well as bulk franckeite XPS spectra). The ratio between Sb and Sn intensities, after normalization to the photoemission cross section and to the microscope transmission, results in an excess of 33% of Sb over Sn. The S 2p core level also indicates two different environments, PbS and $SnS_2$, while Pb 4f has a strong doublet (PbS) and another minor component, possibly due to some lead oxide. In the inset of Fig. 2d we show a low energy electron microscopy (LEEM) image of the flake, appearing as a bright stripe crossing the field of view, with some steps running nearly parallel to one of its edges. From the XPS-PEEM chemical images (inset of Fig. 2e, measured at the Pb $4f_{7/2}$ core level peak), we conclude that franckeite is chemically homogeneous: the image shows a uniform bright stripe through the whole flake, with the same behavior for all the other chemical constituents.

**Liquid phase exfoliation**

Liquid-phase exfoliation (LPE) of layered materials allows scale-up, improves processibility, and opens the way to chemically functionalize the nanosheets in suspension.[22] We demonstrate LPE isolation of franckeite by bath ultrasonication for 1 h at 20 °C, using franckeite as a powder coming from careful grinding mineral pieces (Fig. 3a). First, LPE is carried out in *N*-methyl-2-pyrrolidone (NMP), which has successfully been used to exfoliate graphite and transition metal dichalcogenides,[23, 24] thanks to its surface tension and chemical coordination properties, which favor layer separation and subsequent nanosheet stabilization, respectively. LPE was also investigated in isopropanol/water 1/4 v/v (further referred to as IPA/water), due to matching surface tension with $SnS_2$.[25]

Remarkably, all the suspensions prepared in NMP were indefinitely stable in time (>6 months). Analysis of these colloids using atomic force microscopy (AFM, Figure S8) and TEM (Fig. 3b and c, and S9) reveals large flakes (>100 nm in lateral size) of height <25 nm, together with copious amounts of very small (~10 nm-20 nm in lateral size) fragments.

The LPE exfoliation in IPA/water leads to much more uniform nanosheets. Fig. 3d shows a large area AFM topographic image obtained after drop-casting and drying of a franckeite dispersion prepared in IPA/water on a mica substrate. Statistical analysis of the height data (Fig. 3e) attests to the formation of very thin flakes, with a narrow thickness distribution between 6 and 12 nm. This range corresponds to a maximum number of ~4 to 7 franckeite layers (being one layer an H and Q pair), without taking into account adsorbed solvent molecules remaining onto the nanosheet surface, which might increase the measured thickness by up to 1.2-1.3 nm.[26] The lateral size distribution (Fig. S10), is also more homogeneous and displaced towards larger sheets (c.a. 200 nm). In all cases, the nanosheet composition was ascertained by EDX microanalysis (Fig. S11).

UV-Vis-NIR spectroscopy of the colloidal suspension prepared from the 100 mg·mL$^{-1}$ powder dispersion in NMP was performed after drop-casting and drying of the liquid sample on a glass slide. This allows the elimination of the solvent C–H bonds which result in intense absorption that hinders light transmission measurements in a wide NIR range. The resulting spectrum shows a continuous decrease of the absorption as the wavelength increases from 300 nm to 2750 nm, followed by a wide absorption band from 2750 nm to 3300 nm, reaching a maximum at ~2900 nm (0.43 eV). Together with the absorption onset in the NIR region, the existence of this band is consistent with the narrow bandgap energy determined for franckeite.

The Raman characterization was performed on bulk franckeite powder and on the most concentrated colloidal suspension (from the 100 mg·mL$^{-1}$ NMP powder dispersion). Both samples present similar spectra, with five main bands centered at 66, 145, 194, 253, 318 cm$^{-1}$, and a shoulder from ~400 to 650 cm$^{-1}$ (Fig. 3g), confirming the nature of the colloid obtained. As in a first approximation, franckeite alternates SnS$_2$ and PbS layers having a crystal structure similar to that of SnS$_2$ and PbS (see Fig. 1), most of its Raman bands can be associated to those of the parent structures (see the Supporting Information and Table S2 for a detailed study of the Raman spectra). However, the band at 253 cm$^{-1}$ would result from the combination of phonon modes of both the Q and H layers.[27] We also find differences in the relative intensities between the Raman spectra of bulk and exfoliated material, probably originating from their respective thickness.

Taken together, these results demonstrate franckeite can be efficiently exfoliated both in NMP and IPA/water 1/4, resulting in few-layer nanosheets. The use of NMP leads to more concentrated and stable suspensions, but produces thicker nanosheets and a significant amount of tiny nanoparticles. In contrast, LPE in IPA/water yields more uniform and thinner nanosheets, but the final concentration is significantly lower and the suspensions are less stable.

**Franckeite-based nanodevices**

To further explore the electronic properties of franckeite, we have employed mechanically exfoliated flakes in the fabrication of electronic devices by transferring the flakes onto Ti/Au electrodes pre-patterned on a $SiO_2$/Si substrate. The flakes are placed bridging the electrodes using a deterministic transfer technique with an all dry viscoelastic material.[28] Fig. 4a shows an AFM topographic image of one of these devices (with a thickness ranging from ~7 nm to ~13 nm, see the Supporting Information for optical microscopy characterization). The devices are characterized by measuring current-voltage characteristics (the current passing through the material ($I_{ds}$) while sweeping the drain-source voltage ($V_{ds}$) with a fixed back-gate voltage ($V_g$)) and by measuring the current dependence on the back-gate voltage with a fixed drain-source voltage (transfer curve). Fig. 4b shows the $I_{ds}$-$V_g$ curve for the device shown in Fig. 4a measured in dark conditions, high vacuum (P < $10^{-5}$ mbar) and with an applied $V_{ds}$ = 150 mV. The dependence of the current on the back-gate voltage serves as a test for determining the doping of the material: the decrease in the current with increasing back-gate voltage indicates that the material is p-doped in agreement with the STS and thermopower measurements for bulk. The gate traces also show that the franckeite flake is strongly doped and it cannot be switched off within the experimental gate voltage window. Thus, franckeite does not seem an appropriate material to fabricate FETs and requires doping engineering to reduce the intrinsic doping. This measurement was repeated after 41 days, finding a small drop of 5% in the current intensity, which indicates that the device remains stable over time. This is very relevant when comparing with black phosphorus, the other intrinsically doped p-type 2D semiconductor, which degrades on a time scale of a few hours.[29] From the $I_{ds}$-$V_{ds}$ curve (shown in the inset of Fig 4b) we estimate a resistivity of ~50 mΩ·m.

The absorption spectroscopy of the liquid phase exfoliated material and the STS measurements on bulk franckeite suggest that franckeite presents a narrow bandgap and therefore motivating the application of franckeite as a photodetector working in the NIR (they should be able to generate photocurrents upon illumination with light wavelength as long as ~2000 nm). To test the optoelectronic characteristics of our franckeite-based photodetectors, we first study the dependence of the $I_{ds}$-$V_g$ curves on the illumination with a laser source. The measurements, plotted in Fig. 4c, are carried out using a laser of 640 nm wavelength and show that the drain-source current of the photodetector increases with increasing light power over the full range of the gate voltage.

In order to further characterize the photodetector we calculate the responsivity, a typical figure-of-merit for photodetectors that represents the input-output gain of the device as a function of the laser effective power reaching the device (Fig. 4d). The responsivity ($R$) is calculated as $R = I_{ph} / P_{eff}$, where $I_{ph}$ (photocurrent) is the difference between the current measured upon illumination and in dark conditions, and $P_{eff}$ is the effective power of the laser that reaches the device ($P_{eff} = P_{laser} \cdot A_{device}/A_{spot}$). For the device shown in Fig. 4a, upon illumination at 640 nm, we obtain a maximum responsivity of ~100 mA·W$^{-1}$ for a laser intensity of ~30 mW·cm$^{-2}$. Even if this value is not as high as for other two-dimensional photodetectors, such as monolayer MoS$_2$ ($R > 10^6$ mA·W$^{-1}$) or In$_2$Se$_3$ (R > $10^7$ mA·W$^{-1}$),[30, 31, 32] it is larger than most of the responsivities measured in few-layer black-phosphorus, the two-dimensional phototedector with the narrowest bandgap reported to date,[33, 34] ranging from 0.5 mA·W$^{-1}$ to 135 mA·W$^{-1}$.[35, 36, 37]

We have also studied the photocurrent generation in franckeite photodetectors upon illumination in a wide range of light wavelengths (from 405 nm (UV) to 940 nm (NIR)). The $I_{ds}$-$V_g$ curves, measured in dark conditions and upon illumination with lasers of different wavelengths (Fig. 4e), reveal that the device is able to generate photocurrent at wavelengths as large as 940 nm, in good agreement with the results obtained from the UV-Vis-NIR spectroscopy of the liquid-phase-exfoliated material. The photocurrents calculated from these measurements for two fixed back-gate voltages are plotted in Fig 4f.

## MoS$_2$-franckeite p-n junction

Taking advantage of the simplicity of the deterministic transfer method to build van der Waals heterostructures,[36] and the p-doping of franckeite (uncommon in 2D-materials), we fabricated a p-n junction (building blocks of modern electronics) based on the stacking of mechanically-exfoliated franckeite and MoS$_2$ flakes. Fig. 5a shows an AFM topographic image of the device (see the Supporting Information for optical microscopy characterization), which is represented in an artistic drawing in Fig. 5b: the $MoS_2$ flake is first deposited by deterministic transfer onto a $SiO_2$ substrate in contact with one pre-patterned Ti/Au electrode, then a franckeite flake is deposited in contact with the other pre-patterned electrode, resulting in a van der Waals heterostructure made of a p-doped material (franckeite) and an n-doped material ($MoS_2$) in the overlapping region. The electronic characterization of the device, carried out in vacuum ($P < 10^{-5}$ mbar), at room temperature and in dark conditions, shows diode-like $I_{ds}$-$V_{ds}$ characteristics for different back-gate voltages (Fig. 5c). The $I_{ds}$-$V_g$ curve shown in the inset of Fig. 5c yields a current rectification ratio of 400 and a gate threshold voltage of $V_{th}$ ~ -10 V. To test the optoelectronic properties, the device is illuminated as represented in Fig. 5b with laser spots of 940 nm and 885 nm wavelength, indicating that there is photocurrent generation even for zero applied voltage (short circuit current, $I_{sc}$) and that the current is zero for a finite positive applied voltage (open circuit voltage, $V_{oc}$). This phenomena is due to the photovoltaic effect: upon illumination at zero applied voltage, the photogenerated electron-hole pairs are separated by an internal electric field, generating a photocurrent ($I_{sc}$) with the same sign as the reverse voltage; on the other hand, charge carriers are accumulated at different parts of the device, creating a voltage when the circuit is open ($V_{oc}$) in the forward voltage direction. The photocurrent measured upon illumination with laser spots of 940 nm and 885 nm wavelength presents $I_{sc}$ = -27 pA and $V_{oc}$ = 55 mV at 940 nm, and $I_{sc}$ = -51 pA and $V_{oc}$ = 77 mV at 885 nm (see the Supporting Information for a more detailed analysis of the characteristics of the device). We should stress here that optimizing the performance of franckeite-based p-n junction devices is beyond the scope of this work. Nevertheless, these results demonstrate that one can exploit the p-type character of franckeite in electronic devices where a narrow gap air-stable p-type semiconductor is needed.

**Conclusions**

In summary, we have shown that bulk franckeite can be exfoliated both mechanically and in liquid phase to afford the first naturally-occurring quasi 2D van der Waals heterostructure. The structure and properties of ultrathin flakes of franckeite have been studied extensively from both theory and experiment. Franckeite nanosheets show a very narrow bandgap <0.7 eV and p-type conductivity, and are highly stable under ambient conditions, both as mechanically exfoliated flakes and as colloidal suspensions. These features make it a unique addition to the still rather small library of 2D materials. As validation for its potential technological application, we have constructed prototype photodectectors based on mechanically exfoliated few-layers crystals, as well as a p-n junction made by stacking a $MoS_2$ flake and a franckeite flake.

**Figures**

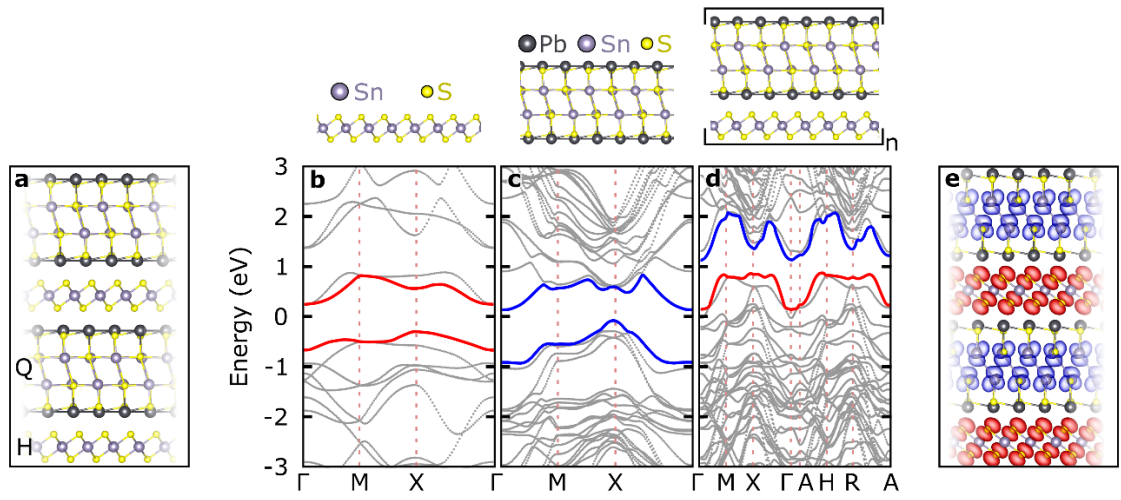

**Figure 1 | Franckeite crystal and band structure. a,** Crystal structure of franckeite: the Q layer includes MX compounds, where M = $Pb^{2+}$ or $Sn^{2+}$ (M can also be $Sb^{3+}$ replacing $Sn^{2+}$) and X = S, while the H layer includes $MX_2$ compounds, where M = $Sn^{4+}$ (M can also be $Fe^{2+}$ replacing $Sn^{4+}$) and X= S. **b,** Calculated band structure for the Q layer that presents a bandgap of ~1 eV. **c,** Calculated band structure for the H layer that presents a bandgap of ~0.7 eV. **d,** Calculated band structure for the franckeite crystal that presents a bandgap of ~0.5 eV. The valence band is given by the Q layer (red line), while the conduction band is given by the H layer (blue line), suggesting that franckeite is a type-II heterostructure. **e,** Bloch states in franckeite in which the valence (red) and conduction (blue) bands are represented.

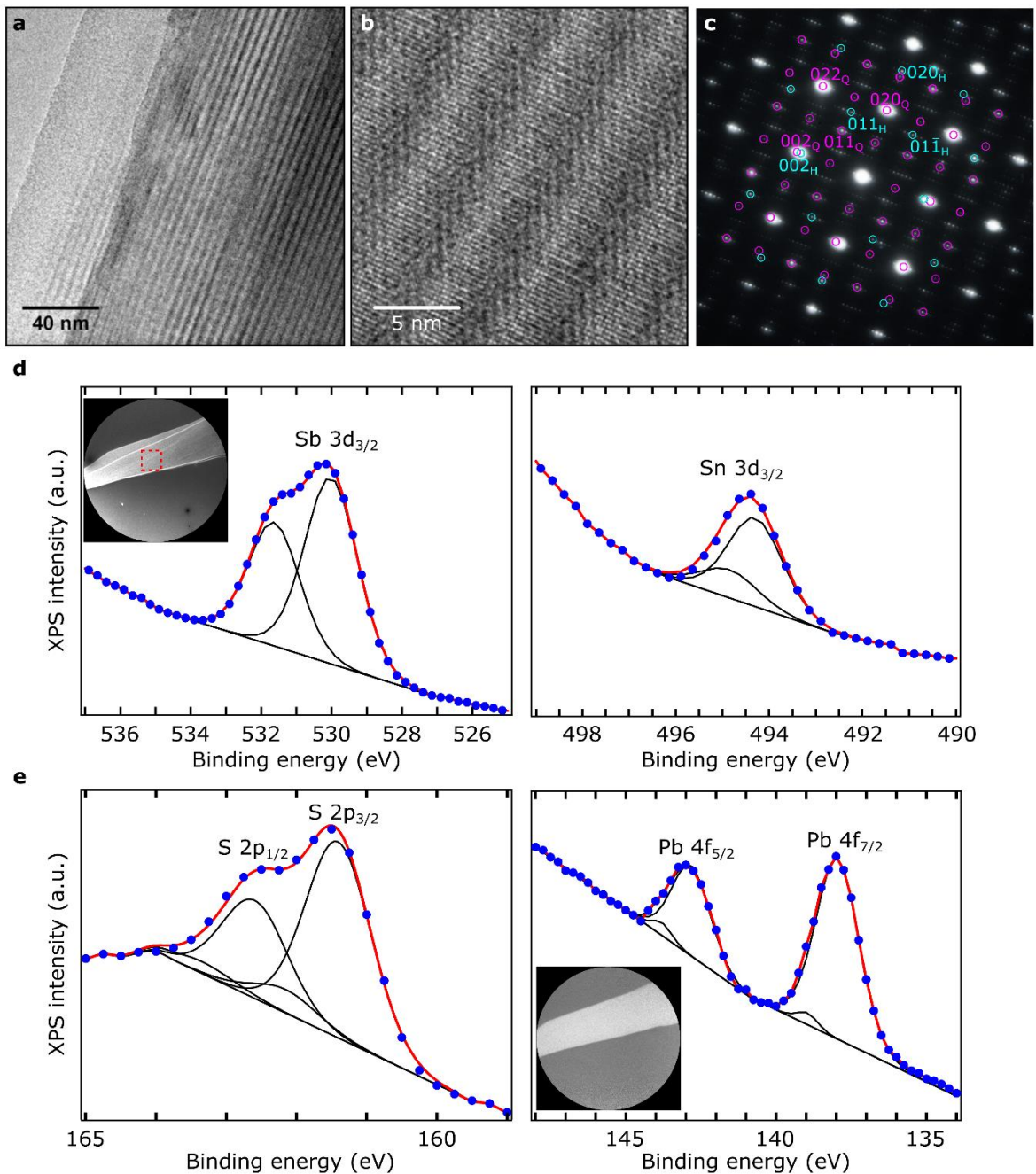

**Figure 2 | Characterization of mechanically exfoliated franckeite flakes. a,** HRTEM micrograph of a franckeite sheet exhibiting the characteristic fringes of franckeite due to the corrugation induced by the misfit between Q and H layers. **b,** Representative atomic scale HRTEM of an ultrathin franckeite layer. **c,** SAED diagram consistent with a misfit layer compound made of PbS and $SnS_2$ layers: Q (purple) and H sublattices (light blue)

lead to the most intense reflections on which superlattice rows of weak intensity are centered. The diagram has been indexed using tetragonal and orthohexagonal vectors for the Q and H phases respectively, according to the orientation and nomenclature defined in Ref. [20]. **d**, Sb $3d_{3/2}$ and Sn $3d_{3/2}$ XPS spectrum acquired with photon energy $hv = 600$ eV. Inset: LEEM image (the field of view is 50 μm and the electron energy is 0.12 eV), the red square indicates the region of integration where the XPS spectra has been acquired. **e,** S $2p_{1/2}$ and $2p_{3/2}$ and Pb $4f_{5/2}$ and $4f_{7/2}$ XPS spectrum acquired with photon energy $hv = 230$ eV. Inset: XPEEM image at Pb $4f_{7/2}$ core level (the field of view is 50 μm and the photon energy is 230 eV). The strong background in the XPS spectra is due to the tail of secondary electrons cascade.

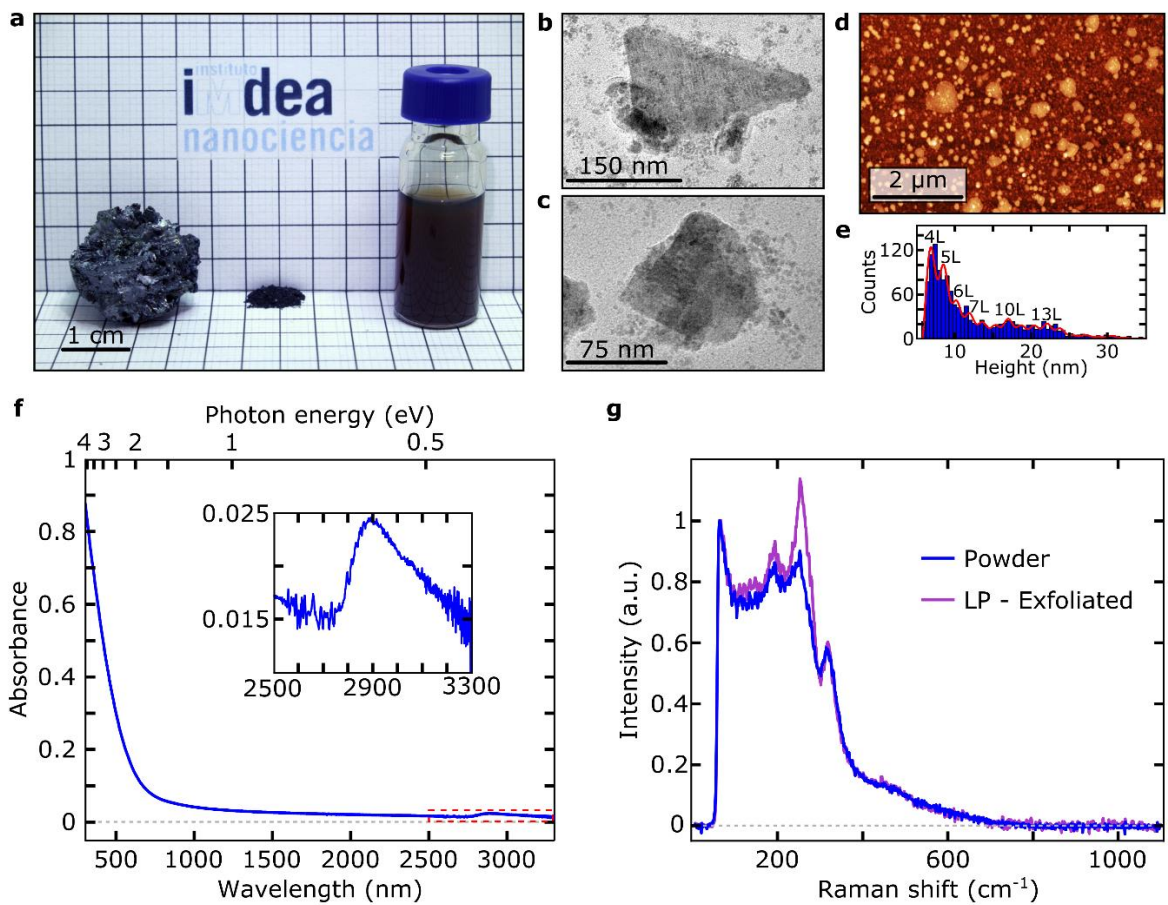

**Figure 3 | Liquid phase exfoliation of franckeite. a**, Franckeite samples. Left: bulk mineral; middle: powder material obtained after grinding of raw chips; right: suspension of exfoliated material prepared by sonication of a 100 mg·mL$^{-1}$ powder dispersion in NMP. **b**, and **c**, TEM images of representative franckeite nanosheets prepared by exfoliation of a 100 mg·mL$^{-1}$ powder dispersion in NMP. **d**, AFM topographic characterization of franckeite nanosheets obtained from the exfoliation of a 1 mg·mL$^{-1}$ powder dispersion in isopropanol/water 1/4 (v/v). **e**, Statistical analysis of the AFM raw height data. The inserted numbers indicate the corresponding number of layers (unit cell, H + Q layer, 1.7 nm in thickness) from ~4 layers (4L) up to ~13 layers (13L). **f**, UV-Vis-NIR spectrum of a thin film of franckeite colloidal suspension deposited on a glass slide; the sample originates

from the liquid-phase exfoliation of a 100 mg·mL$^{-1}$ franckeite powder dispersion. Inset: zoom of the region indicated by a dashed red line that highlights the absorption peak around 2900 nm. **g**, Raman spectra of franckeite raw powder (blue line) and liquid-phase (LP)-exfoliated franckeite obtained from the sonication of a 100 mg·mL$^{-1}$ powder dispersion in NMP (pink line).

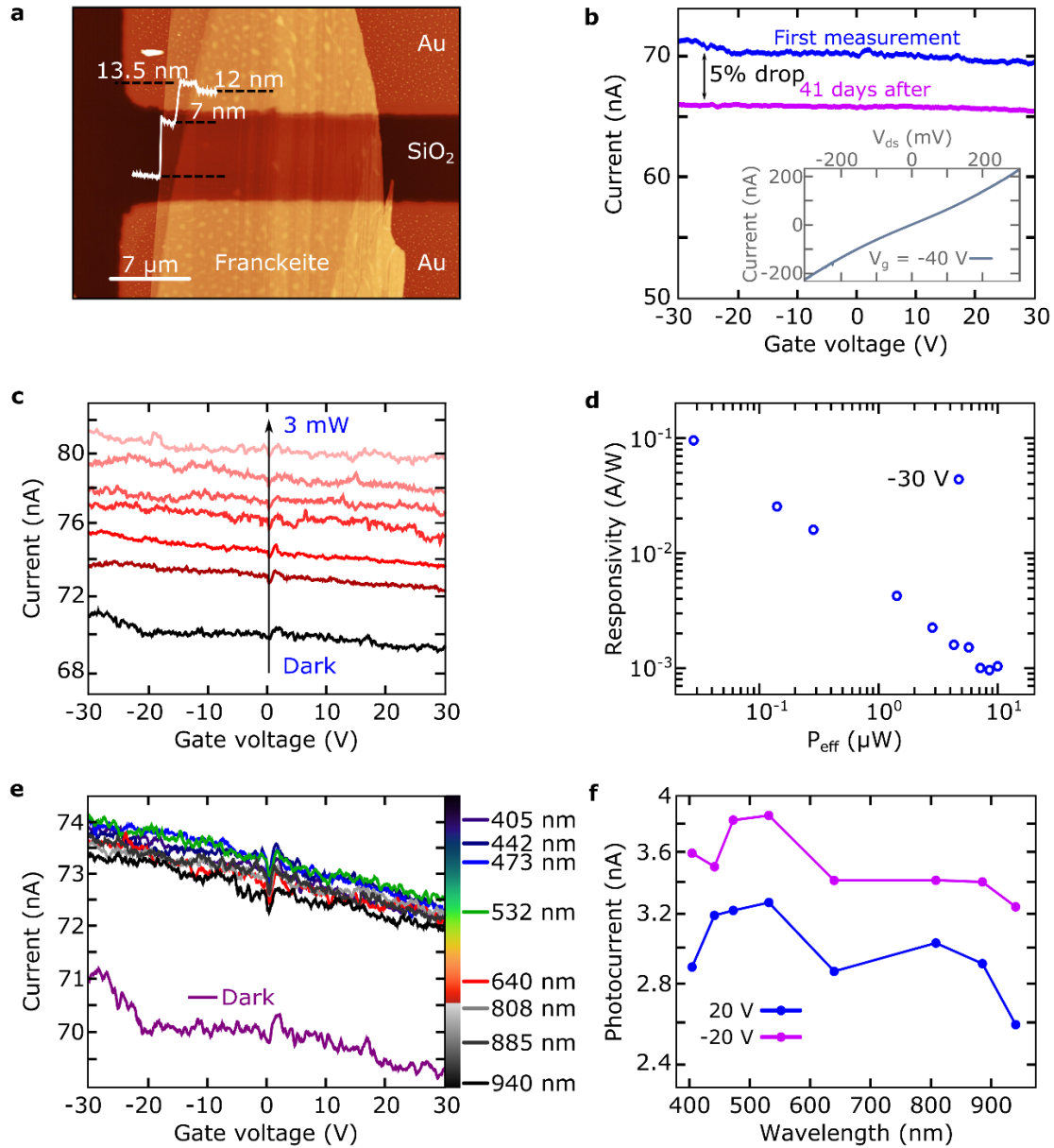

**Figure 4 | Franckeite-based nanodevices. a**, AFM topographic image of a franckeite flake deposited on SiO$_2$ substrate with pre-patterned Ti/Au electrodes. The thickness of the flake ranges from 7 nm (~4 layers) to 13.5 nm (~8 layers). **b**, Current as a function of the applied back-gate voltage in dark conditions for the device shown in **a** ($V_{ds}$ = 150 mV). The gate-dependence shows a p-type doping, hole conduction. The first measurement (blue line) was repeated after 41 days (pink line), showing a drop of 5%, yielding a good stability of the

device. Inset: current-voltage curve with an applied back-gate voltage of -40 V. **c**, Current as a function of the applied back-gate voltage ($V_{ds}$ = 150 mV) for the device shown in **a** in dark conditions and upon illumination with a 640 nm wavelength laser with different powers. **d**, Responsivity of the device shown in a upon illumination with a 640 nm wavelength laser as a function of the laser effective power with an applied back-gate voltage of $V_g$ = -30 V and $V_{ds}$ = 150 mV. **e**, Current as a function of the applied back-gate voltage ($V_{ds}$ = 150 mV) for the device shown in **a** upon ilumination with lasers of different wavelengths at the same intensity ($P_d$ = 6.3 mW·cm$^{-2}$). There is photocurrent generation even at wavelengths as large as 940 nm. **f**, Photocurrent as a function of the laser wavelengths with the same light intensity for back gate voltages of -20 V and +20V and $V_{ds}$ = 150 mV.

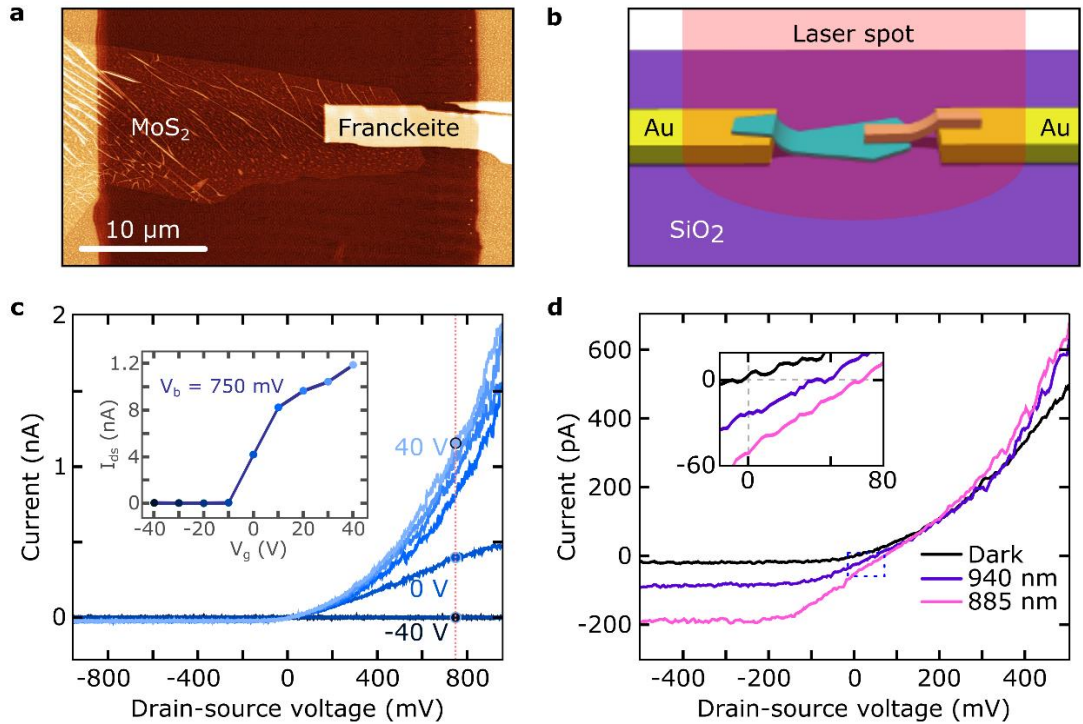

**Figure 5 | p-n junction made by stacking mechanically exfoliated flakes of MoS$_2$ and franckeite**. **a**, AFM topographic image of the p-n junction. **b**, Artistic representation of the p-n junction shown in **a**. **c**, Diode-like current-voltage ($I_{ds}$-$V_{ds}$) curve of the p-n junction in dark conditions for different applied back-gate voltages. Inset: gate trace extracted from the $I_{ds}$-$V_{ds}$ at $V_{ds}$ = 750 mV. The p-n junction switches on at an applied back-gate voltage of 0 V. **d**, Diode-like current-voltage ($I_{ds}$-$V_{ds}$) curve of the p-n junction at an applied back-gate voltage of $V_g$ = 40 V in dark conditions and upon illumination with laser of 940 nm and 885 nm wavelength, both with a power of 140 μW. The inset highlights the region around $V_{ds}$ = 0 V and $I_{ds}$ = 0 V to show the short-circuit current ($I_{sc}$) and open circuit voltage ($V_{oc}$) values, obtaining $I_{sc}$ = -27 pA and $V_{oc}$ = 55 mV at 940 nm, and $I_{sc}$ = -51 pA and $V_{oc}$ = 77 mV at 885 nm.

**Methods**

**Materials:** bulk franckeite mineral from mine San José, Oruro (Bolivia) was used for both mechanical and chemical exfoliation.

**Density Functional Theory calculations:** calculations of the electronic properties of franckeite are based on the framework of density functional theory (DFT), as implemented in the Quantum ESPRESSO package.[18] The generalized gradient approximation of Perdew-Burke-Ernzerhof (GGA-PBE) was adopted for exchange-correlation functional.[38] The electron-ion interaction is described using the norm-conserving Troullier-Martins pseudopotentials are employed in PBE calculations.[39] The energy cut-off for the plane wave basis set is put at 60 Ry with a charge density cut-off of 240 Ry. We have used a Monkhorst-Pack scheme with a 5×5×3 k-mesh for the Brillouin zone integration for the supercell (22 atoms).[40] Herein, we employed the van der Waals interaction described within a semiempirical approach following the Grimme formula.[41]

**Electron microscopies and X-ray photoelectron spectroscopy of mechanically exfoliated flakes:**

*HRTEM, SAED and SEM.* Mechanically-exfoliated franckeite layers were transferred onto a holey $Si_3N_4$ membrane window grid and characterized using a JEOL JEM 3000F microscope (TEM, 300 kV) and observed with a Zeiss EVO MA15 microscope (SEM, 15 kV).

*XPS*. Low energy electron microscopy and micro-XPS measurements were done at the LEEM microscope (Elmitec, GmbH) in operation at ALBA synchrotron (Barcelona, Spain).[21] The instrument is equipped with a LaB6 electron gun for real space imaging (LEEM), with lateral resolution of 20 nm. The imaging column of the microscope has an electron energy filter, and using the photons of tunable energy coming from the synchrotron (at 16º incidence angle), it is possible to perform X-ray photoemission spectroscopy with an energy resolution of 0.25 eV keeping the lateral resolution.

**Liquid-phase exfoliation:**

*Colloid preparation.* Fragments of cleaved natural franckeite were grinded in a porcelain mortar until making a finely grained and black powder. This powder was dispersed in NMP (5 mL) at various concentrations (0.1, 1, 10 and 100 mg·mL$^{-1}$) in 20 mL glass vials. Each dispersion was sonicated for 1 h in a Fisher Scientific FB 15051 ultrasonic bath (37 kHz, 280 W, ultrasonic peak max. 320 W, standard sine-wave modulation) thermostated at 20 °C. Resulting black suspensions were centrifuged (990 *g*, 30 min, 25 °C, Beckman Coulter Allegra® X-15R, FX6100 rotor, radius 9.8 cm) to remove poorly exfoliated solid. After centrifugation, the corresponding supernatants were carefully isolated from the black sediments to obtain exfoliated franckeite pale-to-dark orange colloidal suspensions respectively, stable over at least 6 months. This NMP exfoliation proved to be highly reproducible.

Exfoliation of 1 mg·mL$^{-1}$ samples was carried out in similar conditions in an IPA/water mixture (1/4, v/v) and resulted in pale orange suspensions that precipitated within 24 h. The experiment was found more difficult to replicate, sometimes leading to particle-free supernatants after centrifugation. Nevertheless, redispersion of corresponding black sediments in NMP (5 mL, without sonication) and subsequent centrifugation (990 $g$, 30 min, 25 °C) consistently afforded stable exfoliated franckeite colloidal suspensions (same absorption range as successfully prepared IPA/water samples). So we attribute the poor reproducibility in IPA/water to the weaker dispersing abilities of this solvent.

*AFM.* Colloidal suspensions were drop-casted on freshly cleaved mica substrates and dried under vacuum. Images were acquired using a JPK NanoWizard II AFM working in tapping (NMP) or contact (IPA/water) mode at room temperature in air.

*TEM*. Colloidal suspensions were drop-casted onto 200 square mesh copper grids covered with a carbon film and observed using a JEOL JEM 2100 microscope (200 kV), equipped with an Energy Dispersive X-Ray (EDX) detector.

*UV-Vis-NIR spectroscopy.* As-prepared colloidal suspension was drop-casted onto a microscope glass slide and dried at 60°C. The sequence was repeated until a pale orange thin film appeared whose absorption was characterized using a Cary 5000 spectrophotometer, Agilent Technologies (wavelength range: 175-3300 nm). The measurement was replicated three times with different glass slides.

*Raman spectroscopy.* Powder (pressed) and liquid-phase-exfoliated franckeite (dried at room temperature) deposited on glass slides were analyzed with a WITec alpha300 RA combined confocal Raman imaging/atomic force microscope (objective NA 0.95, 100×; laser excitation: 532 nm, 0.2 mW). Powder- and exfoliated-material spectra result from the average of 200 and 5 measurements over the respective samples.

**Franckeite-based nanodevices:**

*Sample fabrication.* A franckeite flake is exfoliated from the parent crystal with a polydimethylsiloxane (PDMS) stamp (Gelfilm® from Gelpak). Then, the PDMS stamp with the desired flake is brought into contact with the two Au electrodes on $SiO_2$ and peeled off carefully, leaving the flake on the substrate. In this way we obtain three-electrodes devices (details about deterministic transfer can be found in Ref.[28]).

*AFM.* AFM characterization of the devices was carried out using a Digital Instruments D3100 AFM operated in the amplitude modulation mode.

*Electronic and optoelectronic characterization.* The electronic and optoelectronic characterization have been performed in a Lakeshore Cryogenics tabletop probe station at room temperature and high vacuum ($<10^5$ mbar). The light excitation is provided by diode pump solid state lasers operated in continuous mode and guided with an optical fiber which yields a spot of 200 µm in diameter on the sample.

**Acknowledgements**

A.C-G. acknowledges financial support from the BBVA Foundation through the fellowship "I Convocatoria de Ayudas Fundacion BBVA a Investigadores, Innovadores y Creadores Culturales" ("Semiconductores ultradelgados: hacia la optpelectronica flexible"), from the MINECO (Ramón y Cajal 2014 program, RYC-2014-01406) and from the MICINN (MAT2014-58399-JIN). A.J.M-M., G.R-B. and N.A. acknowledge the support of the MICCINN/MINECO (Spain) through the programs MAT2014-57915-R, BES-2012-057346 and FIS2011-23488 and Comunidad de Madrid (Spain) through the program S2013/MIT-3007 (MAD2D). J.I. and H.S.J.vd.Z acknowledge the support of the Dutch organization for Fundamental Research on Matter (FOM) and by the Ministry of Education, Culture, and Science (OCW). M.A.N.O. acknowledeges the support of the MICCINN/MINECO (Spain) through the programs MAT2013-49893-EXP and MAT2014-59315-R. Authors M.A.N., A.J.M-M. and A.C-G. acknowledge the support from ALBA Synchrotron for the experiments performed at Circe beamline (BL24-CIRCE) at ALBA Synchrotron with the collaboration of ALBA staff (proposal ID 2015091399). W. S. P. acknowledges CAPES Foundation, Ministry of Education of Brazil, under grant BEX 9476/13-0. WSP and JJP acknowledge MICCINN/MINECO (Spain) for financial support under grant FIS2013-47328-C02-1; the European Union structural funds and the Comunidad de Madrid MAD2D-CM program under grant nos. P2013/MIT-3007 and P2013/MIT-2850; the Generalitat Valenciana under grant no. PROMETEO/2012/011. WSP and JJP also acknowledge the computer resources and assistance provided by the Centro de Computación Científica of the Universidad Autónoma de Madrid and the RES.

**Author contributions**

A.J.M-M. and J.O.I. performed the optoelectronic characterization of the photodetectors and the p-n junction, fabricated by A.C-G. E.G. performed the SEM, TEM, SAED characterization of the material and the liquid-phase exfoliation with the corresponding characterization. W.P. performed the DFT calculations of the band structure of franckeite.

**Additional information**

Supporting Information is available in the online version of the paper. Reprints and permissions information is available online at www.nature.com/reprints. In the Supporting Information we include density functional theory calculations of Sb-doped franckeite, scanning tunneling microscopy characterization, scanning electron microscopy characterization, XPS on franckeite thick chips, liquid-phase exfoliation, powder and liquid-phase exfoliated franckeite Raman spectra interpretation, thermopower, optical microscopy characterization of a franckeite photodetector and $MoS_2$-franckeite p-n junction.

**Competing financial interests**

The authors declare no competing financial interests.

# References

1. Dean CR, Young AF, MericI, LeeC, WangL, SorgenfreiS, *et al.* Boron nitride substrates for high-quality graphene electronics. *Nat Nano* 2010, **5**(10)**:** 722-726.

2. Geim AK, Grigorieva IV. Van der Waals heterostructures. *Nature* 2013, **499**(7459)**:** 419-425.

3. Bae S, Kim H, Lee Y, Xu X, Park J-S, Zheng Y, *et al.* Roll-to-roll production of 30-inch graphene films for transparent electrodes. *Nat Nano* 2010, **5**(8)**:** 574-578.

4. Yang W, Chen G, Shi Z, Liu C-C, Zhang L, Xie G, *et al.* Epitaxial growth of single-domain graphene on hexagonal boron nitride. *Nat Mater* 2013, **12**(9)**:** 792-797.

5. Gong Y, Lin J, Wang X, Shi G, Lei S, Lin Z, *et al.* Vertical and in-plane heterostructures from WS2/MoS2 monolayers. *Nat Mater* 2014, **13**(12)**:** 1135-1142.

6. Zhang X, Meng F, Christianson JR, Arroyo-Torres C, Lukowski MA, Liang D, *et al.* Vertical Heterostructures of Layered Metal Chalcogenides by van der Waals Epitaxy. *Nano Letters* 2014, **14**(6)**:** 3047-3054.

7. Ponomarenko LA, Geim AK, Zhukov AA, Jalil R, Morozov SV, Novoselov KS, *et al.* Tunable metal-insulator transition in double-layer graphene heterostructures. *Nat Phys* 2011, **7**(12)**:** 958-961.

8. Georgiou T, Jalil R, Belle BD, Britnell L, Gorbachev RV, Morozov SV, *et al.* Vertical field-effect transistor based on graphene-WS2 heterostructures for flexible and transparent electronics. *Nat Nano* 2013, **8**(2)**:** 100-103.

9. Bertolazzi S, Krasnozhon D, Kis A. Nonvolatile Memory Cells Based on MoS2/Graphene Heterostructures. *ACS Nano* 2013, **7**(4)**:** 3246-3252.

10. Hunt B, Sanchez-Yamagishi JD, Young AF, Yankowitz M, LeRoy BJ, Watanabe K, *et al.* Massive Dirac Fermions and Hofstadter Butterfly in a van der Waals Heterostructure. *Science* 2013, **340**(6139)**:** 1427-1430.

11. Withers F, Del Pozo-Zamudio O, Mishchenko A, Rooney AP, Gholinia A, Watanabe K, *et al.* Light-emitting diodes by band-structure engineering in van der Waals heterostructures. *Nat Mater* 2015, **14**(3)**:** 301-306.

12. Ross JS, Klement P, Jones AM, Ghimire NJ, Yan J, Mandrus DG, *et al.* Electrically tunable excitonic light-emitting diodes based on monolayer WSe2 p-n junctions. *Nat Nano* 2014, **9**(4)**:** 268-272.

13. Baugher BWH, Churchill HOH, Yang Y, Jarillo-Herrero P. Optoelectronic devices based on electrically tunable p-n diodes in a monolayer dichalcogenide. *Nat Nano* 2014, **9**(4)**:** 262-267.

14. Li L, Yu Y, Ye GJ, Ge Q, Ou X, Wu H, *et al.* Black phosphorus field-effect transistors. *Nat Nano* 2014, **9**(5)**:** 372-377.

15. Castellanos-Gomez A, Vicarelli L, Prada E, Island JO, Narasimha-Acharya KL, Blanter SI, *et al.* Isolation and characterization of few-layer black phosphorus. *2D Materials* 2014, **1**(2)**:** 025001.

16. Liu H, Neal AT, Zhu Z, Luo Z, Xu X, Tománek D, *et al.* Phosphorene: An Unexplored 2D Semiconductor with a High Hole Mobility. *ACS Nano* 2014, **8**(4)**:** 4033-4041.

17. Makovicky E, Petříček V, Dušek M, Topa D. The crystal structure of franckeite, $Pb_{21.7}Sn_{9.3}Fe_{4.0}Sb_{8.1}S_{56.9}$. *American Mineralogist* 2011, **96**(11-12)**:** 1686-1702.

18. Giannozzi P, Baroni S, Bonini N, Calandra M, Car R, Cavazzoni C, *et al.* QUANTUM ESPRESSO: a modular and open-source software project for quantum simulations of materials. *J Phys: Condens Matter* 2009, **21:** 395502.

19. Terrones H, López-Urías F, Terrones M. Novel hetero-layered materials with tunable direct band gaps by sandwiching different metal disulfides and diselenides. *Scientific Reports* 2013, **3:** 1549.

20. Williams TB, Hyde BG. Electron microscopy of cylindrite and franckeite. *Physics and Chemistry of Minerals* 1988, **15**(6)**:** 521-544.

21. Aballe L, Foerster M, Pellegrin E, Nicolas J, Ferrer S. The ALBA spectroscopic LEEM-PEEM experimental station: layout and performance. *Journal of synchrotron radiation* 2015, **22**(3)**:** 0-0.

22. Nicolosi V, Chhowalla M, Kanatzidis MG, Strano MS, Coleman JN. Liquid Exfoliation of Layered Materials. *Science* 2013, **340**(6139).

23. Hernandez Y, Nicolosi V, Lotya M, Blighe FM, Sun Z, De S, et al. High-yield production of graphene by liquid-phase exfoliation of graphite. *Nature Nanotechnology* 2008, **3**(9): 563-568.

24. Coleman JN, Lotya M, O'Neill A, Bergin SD, King PJ, Khan U, et al. Two-Dimensional Nanosheets Produced by Liquid Exfoliation of Layered Materials. *Science* 2011, **331**(6017): 568-571.

25. Shen J, He Y, Wu J, Gao C, Keyshar K, Zhang X, et al. Liquid Phase Exfoliation of Two-Dimensional Materials by Directly Probing and Matching Surface Tension Components. *Nano Letters* 2015, **15**(8): 5449-5454.

26. Zeng H, Cui X. An optical spectroscopic study on two-dimensional group-VI transition metal dichalcogenides. *Chemical Society Reviews* 2015, **44**(9): 2629-2642.

27. Ovsyannikov SV, Shchennikov VV, Cantarero A, Cros A, Titov AN. Raman spectra of $(PbS)_{1.18}(TiS_2)_2$ misfit compound. *Mater Sci Eng A* 2007, **462**(1–2): 422-426.

28. Castellanos-Gomez A, Buscema M, Molenaar R, Singh V, Janssen L, van der Zant HSJ, et al. Deterministic transfer of two-dimensional materials by all-dry viscoelastic stamping. *2D Materials* 2014, **1**(1): 011002 (011008 pp.)-011002 (011008 pp.).

29. Joshua OI, Gary AS, Herre SJvdZ, Andres C-G. Environmental instability of few-layer black phosphorus. *2D Materials* 2015, **2**(1): 011002.

30. Lopez-Sanchez O, Lembke D, Kayci M, Radenovic A, Kis A. Ultrasensitive photodetectors based on monolayer MoS2. *Nat Nano* 2013, **8**(7): 497-501.

31. Zhang W, Huang J-K, Chen C-H, Chang Y-H, Cheng Y-J, Li L-J. High-Gain Phototransistors Based on a CVD MoS2 Monolayer. *Advanced Materials* 2013, **25**(25): 3456-3461.

32. Island JO, Blanter SI, Buscema M, van der Zant HSJ, Castellanos-Gomez A. Gate Controlled Photocurrent Generation Mechanisms in High-Gain In2Se3 Phototransistors. *Nano Lett* 2015, **15**(12): 7853-7858.

33. Engel M, Steiner M, Avouris P. Black Phosphorus Photodetector for Multispectral, High-Resolution Imaging. *Nano Letters* 2014, **14**(11): 6414-6417.

34. Castellanos-Gomez A. Black Phosphorus: Narrow Gap, Wide Applications. *The Journal of Physical Chemistry Letters* 2015, **6**(21): 4280-4291.

35. Buscema M, Groenendijk DJ, Blanter SI, Steele GA, van der Zant HSJ, Castellanos-Gomez A. Fast and Broadband Photoresponse of Few-Layer Black Phosphorus Field-Effect Transistors. *Nano Lett* 2014, **14**(6)**:** 3347-3352.

36. Buscema M, Groenendijk DJ, Steele GA, van der Zant HSJ, Castellanos-Gomez A. Photovoltaic effect in few-layer black phosphorus PN junctions defined by local electrostatic gating. *Nat Commun* 2014, **5**.

37. Youngblood N, Chen C, Koester SJ, Li M. Waveguide-integrated black phosphorus photodetector with high responsivity and low dark current. *Nat Photon* 2015, **9**(4)**:** 247-252.

38. Perdew JP, Burke K, Ernzerhof M. Generalized Gradient Approximation Made Simple. *Physical Review Letters* 1996, **77**(18)**:** 3865-3868.

39. Monkhorst HJ, Pack JD. Special points for Brillouin-zone integrations. *Physical Review B* 1976, **13**(12)**:** 5188-5192.

40. Troullier N, Martins JL. Efficient pseudopotentials for plane-wave calculations. *Physical Review B* 1991, **43**(3)**:** 1993-2006.

41. Grimme S. Semiempirical GGA-type density functional constructed with a long-range dispersion correction. *Journal of Computational Chemistry* 2006, **27**(15)**:** 1787-1799.

**Additional information**

# Franckeite: a naturally occurring van der Waals heterostructure

*Aday J. Molina-Mendoza,[1,†] Emerson Giovanelli,[2,†] Wendel S. Paz,[1] Miguel Angel Niño,[2] Joshua O. Island,[3] Charalambos Evangeli,[1,§] Lucía Aballe,[4] Michael Foerster,[4] Herre S. J. van der Zant,[3] Gabino Rubio-Bollinger,[1,5] Nicolás Agraït,[1,2,5] J. J. Palacios,[1,*] Emilio M. Pérez,[2,*] and Andres Castellanos-Gomez.[2,*]*

[1]Departamento de Física de la Materia Condensada, Universidad Autónoma de Madrid, Campus de Cantoblanco, E-28049, Madrid, Spain.

[2]Instituto Madrileño de Estudios Avanzados en Nanociencia (IMDEA-Nanociencia), Campus de Cantoblanco, E-28049 Madrid, Spain.

[3]Kavli Institute of Nanoscience, Delft University of Technology, Lorentzweg 1, 2628 CJ Delft, The Netherlands.

[4]ALBA Synchrotron Light Facility, Carrer de la Llum 2-26, Cerdanyola del Vallés, Barcelona 08290, Spain.

[5]Condensed Matter Physics Center (IFIMAC), Universidad Autónoma de Madrid, E-28049 Madrid, Spain.

[†]*These authors contributed equally.*

§*Present address:* Department of Physics, Lancaster University, Lancaster LA1 4YB, United Kingdom.

*email: juanjose.palacios@uam.es; emilio.perez@imdea.org; andres.castellanos@imdea.org.

In this Supporting Information we include:

- Density functional theory calculations of Sb-doped franckeite
- Scanning tunneling microscopy characterization
- Scanning electron microscopy characterization
- XPS on franckeite mineral chips
- Liquid phase exfoliation
- Powder and liquid-phase-exfoliated franckeite Raman spectra interpretation
- Thermopower
- Optical microscopy characterization of a franckeite photodetector
- $MoS_2$-franckeite p-n junction

**Density functional theory calculations of Sb-doped franckeite**

In Fig. 1 of the main text we show the density functional theory calculations of the band structure of franckeite considering that the crystal is composed of only PbS and $SnS_2$ layers. Nevertheless, it is found that the composition of franckeite includes Sb substitutional atoms replacing Sn atoms which may donate one more electron to the system as a cation in the H layer. We compute the calculations for a 50% and 100% substitution, i.e., 50% (or 100%) of the Sn atoms have been replaced by Sb atoms in the H layer. In Fig. S1a we show the calculated band structure of franckeite composed only of PbS and $SnS_2$ (it is also shown in Fig. 1d of the main text), where we find a bandgap of ~ 0.4 eV with the Fermi level lying below the valence band. When we introduce an Sb substitution of 50% of (Fig. S1b), one of the valence bands becomes almost fully occupied, while for a 100% substitution (Fig. S1c) both bands are essentially full, leaving a small concentration of holes in the system. The gap has been closed at this stage, but this can be attributed to the known deficiencies of the standard approximation to the density functional that we have used and we expect it to remain open in reality. Substitutional Sb in the Q slab or substitutional Fe in the H layer, as well as an increase in the percentage of Pb in the Q slab, does not fundamentally change the spatial separation of electrons and holes in the heterostructure. The saturation with Sb in the H layer, maybe beyond nominal concentration values, seems to be determinant for the semiconducting behavior to emerge, but this coincides with the strong presence of $Sb^{4+}$ observed in our samples by micro-XPS and EDX.

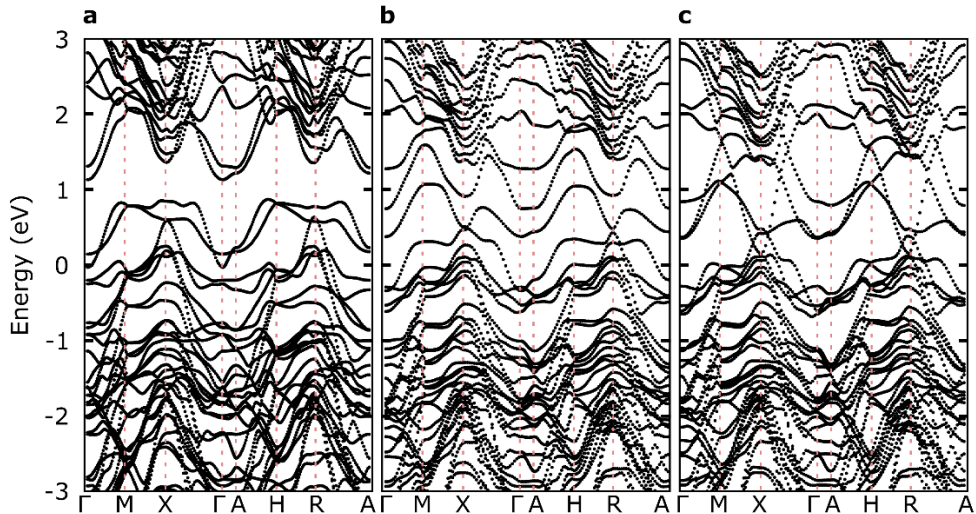

**Figure S1 | Franckeite band structure with Sb doping in the H layer. a,** Calculated band structure of franckeite without Sb. **b,** Calculated band structure of franckeite when 50% of the Sn atoms have been replaced by Sb atoms in the H layer. **c,** Calculated band structure of franckeite when 100% of the Sn atoms have been replaced by Sb atoms in the H layer.

Now we proceed to investigate the case in which the substitutional Sb atoms appear in both the Q and the H layer. We fix the amount of Sb in the Q layer at a substitution of 50% of the Sn atoms, while we change the amount of Sb from 0% (Fig. S2a), 50% (Fig. S2b) and 100% (Fig. S2c) in the H layer. Here we observe that the amount of Sb doping is crucial to open a gap in the band structure and, in the situation of 50% Sb, 50% Sn in the Q layer and 100% Sb in the H layer, an indirect bandgap of ~ 0.4 eV appears. These results confirm that the present of Sb in the franckeite composition is changing the band structure in such a way that it opens a small bandgap and provides a p-doping, although these calculations are just approaches to the real franckeite structure.

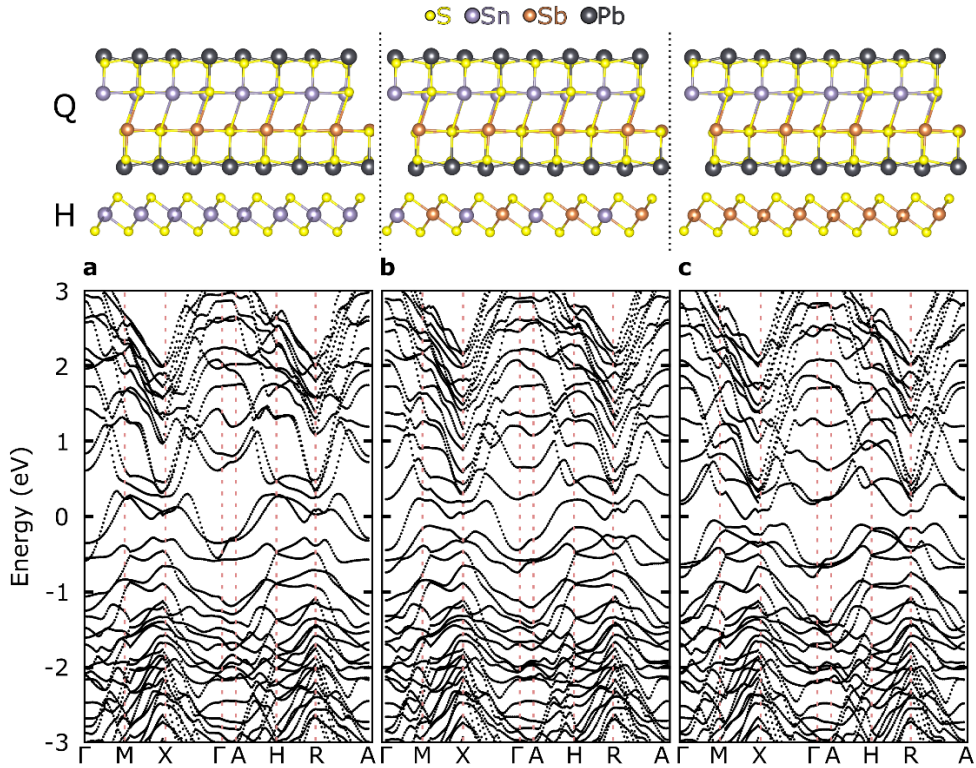

**Figure S2 | Franckeite band structure with Sb doping in the H and Q layers. a,** Calculated band structure of franckeite with 50% Sn and 50% Sb in the Q layer and 0% Sb in the H layer. **b,** Calculated band structure of franckeite with 50% Sn and 50% Sb in the Q layer and 50% Sn and 50% Sb in the H layer. **c,** Calculated band structure of franckeite with 50% Sn and 50% Sb in the Q layer and 100% Sb in the H layer.

**Scanning tunneling microscopy characterization**

Bulk franckeite material was investigated by means of scanning tunneling microscopy (STM) and spectroscopy (STS). For this purpose, a piece of franckeite (Fig. S3a) is placed on a Au substrate which serves as electrode (indium is used to ensured electrical contact between the two of them), and a mechanically cut Au wire is used as a tip. Franckeite is then exfoliated to obtain a clean surface. Fig. S3b shows a STM topographic image of the franckeite surface taken with an applied bias voltage of $V_{bias} = 1$ V and at room temperature and ambient pressure, where a step-edge can be identified with a height of ~ 2.3 nm (~ 1 or 2 layers, depending on if the top layer is a $SnS_2$ or a PbS layer). The average of a set of 200 STS current-voltage characteristics is shown in Fig. S3c, depicting a clear semiconducting *p-doped* behavior: there is a zero conductance zone around zero bias voltage which extends between two well-defined values, corresponding to the valence band ($B_V$) and conduction band ($B_C$) of the material; the doping of the material can be determined from the fact that the curve is "shifted to the left", meaning that the Fermi level (corresponding to the zero bias voltage) is closer to the valence band than to the conduction band. The whole set of 200 current-voltage curves is shown in the inset in logarithmic scale where the $B_V$ and $B_C$ values can be extracted, obtaining $B_V = -0.25$ eV and $B_C = 0.35$ eV, yielding and electronic bandgap of $E_{g,el} = 0.6$ eV. At this point, the thermal broadening of the bands due to the temperature of the sample must be considered in the calculation, adding Et ≈ $3.5k_BT = 90$ meV at room temperature,[1] therefore, the electronic bandgap of franckeite is $E_{g,el} \approx 0.7$ eV.

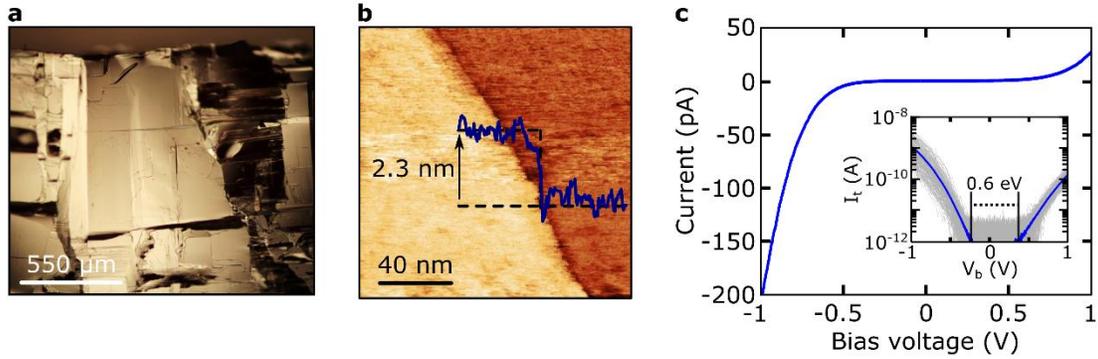

**Figure S3 a**, Optical microscopy photograph of a piece of franckeite. The material presents large area flat terraces after mechanical exfoliation. **b,** STM topographic image of bulk franckeite. **c**, STS characterization of franckeite, here we show the average curve obtained from a set of 200 current-voltage curves, which clearly shows a semiconducting p-doped since the zero bias voltage (coinciding with the Fermi level) is closer to the valence band than to the conduction band. Inset: STS current-voltage curves in logarithmic scale, the valence and conduction bands values (-0.25 eV and 0.35 eV, respectively) are highlighted with arrows, yielding an electronic bandgap of ~ 0.6 eV (~0.7 eV after correction of the thermal broadening).

**Scanning electron microscopy characterization**

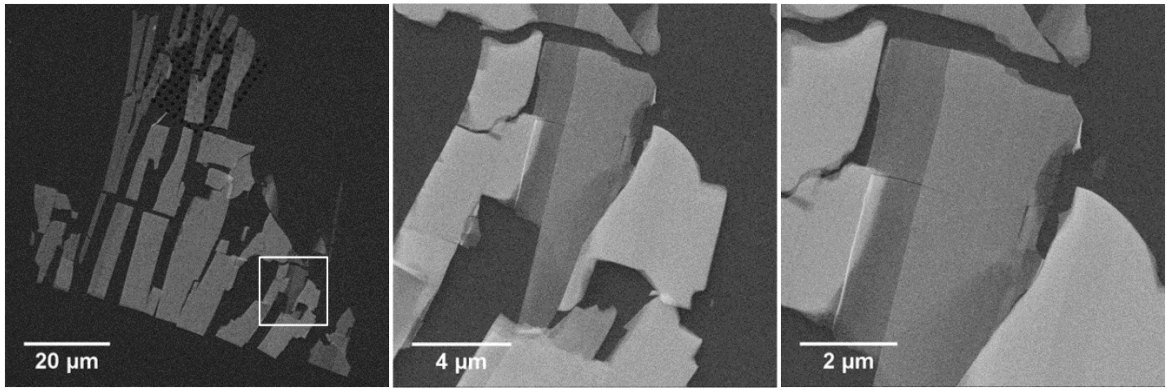

**Figure S4 | Scanning Electron Microscopy images of mechanically exfoliated franckeite. a,** SEM image of several ultrathin franckeite layers deposited on a TEM grid. **b,** Zoom in on the white squared area marked in a, showing layers of various thicknesses. Conditions: 15 kV, 11 pA, working distance 6.5 mm. **c,** Zoom in on the image shown in b.

**XPS on franckeite mineral chips**

In addition to the local spectroscopy on a single flake in the main text we have performed the chemical characterization of the franckeite mineral with standard XPS, using the Al Kα line from a conventional monochromatized X-ray source described elsewhere[2]: several fragments of a few millimeters in size of mineral chips covering a platinum surface were analyzed. There are no photoemission data in the literature about franckeite, but the overview photoemission spectrum (not shown) of the mineral gives the atomic components and contaminants of the franckeite surface: Sb, Sn, Pb and S, as well as the other typical minor components of franckeite Fe and Ag,[3, 4, 5] appearing as contaminants oxygen and carbon. The XPS narrow scans give information on the oxidation state and the possible adscription to the different structural layers of the franckeite.

In the minor components of franckeite, iron is present in a very small concentration, although it is apparent in the Sn $3p_{3/2}$ narrow scan (Fig. S5a). The broad shape and the position of Fe $2p_{3/2}$ at 708 eV in binding energy, and the distance of 13.8 eV from Fe $2p_{1/2}$, suggest $Fe^{2+}$ or $Fe^{3+}$ valence state similar to some iron sulfides ($Fe_7S_8$, $FeS_2$). Silver is typical in cylindrite group of minerals like incaite and potosiite, and it is present in this sample in a small amount too: Ag $3d_{5/2}$ core level (Fig. S5b) appears at 367.8 eV binding energy, that could be compatible with Ag inclusions in an oxide state. As contaminants we find carbon and oxygen: C 1s peak at 285.0 eV; oxygen core level O1s is not directly observable due to its coincidence with Sb 3d that it is one of the most intense peaks, but the oxygen Auger KLL transition is visible in the overview spectrum. After fitting the components for Sb 3d core level (Fig. S6) the O1s appear under Sb 3d5/2 as a broad (fwhm of 2.75 eV) peak at 531.5 eV, indicating several oxygen states.

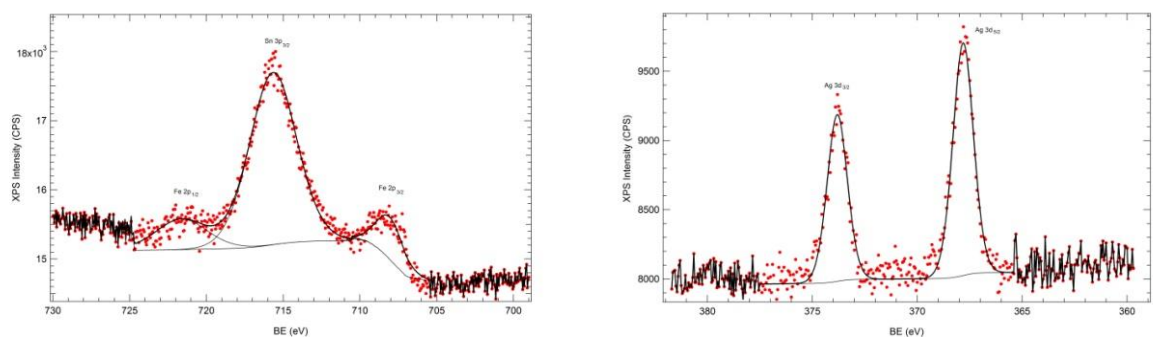

**Figure S5.** XPS spectra of (a) Fe 2p region, coincident with Sn $3p_{3/2}$ core level and (b) Ag 3d region.

XPS narrow spectra of Sb 3d, Sn 3d, S 2p and Pb 4f regions of mineral franckeite are shown in Fig. S6. The fits to these spectra are generally satisfactory, with the exception of the poor fit at the low binding energy tail on the Sb 3d and Sn 3d spectra. The corresponding binding energies extracted from the fits are shown in Table S1 and compared to the binding energies of the same components for the single flake experiment measured with synchrotron radiation technique. In the last case we measured a shift towards higher binding energy of 0.3 – 0.4 eV for the low binding energy components (S and Pb measured at $hv = 230$ eV), and 0.6 – 1.0 eV for Sb (measured at $hv = 600$ eV), with a good agreement for the case of Sn. As we do not expect different chemical species in both cases, because the flake comes from the same mineral fragment, we attribute these shifts towards higher binding energies to a charging effect due to the much higher photon flux in the synchrotron radiation experiment than in the conventional Al Kα x ray source (depending on the energy shift in the particular photon flux).

|  | Sb $3d_{3/2}$ | Sb $3d_{5/2}$ | Sn $3d_{3/2}$ | Sn $3d_{5/2}$ | S $2p_{1/2}$ | S $2p_{3/2}$ | Pb $4f_{5/2}$ | Pb $4f_{7/2}$ |
|---|---|---|---|---|---|---|---|---|
| A-doublet $_{Min}$ | 538.7 | 529.4 | 494.3 | 485.9 | 162.2 | 161.0 | 142.60 | 137.7 |
| B-doublet $_{Min}$ | 539.8 | 530.4 | 495.2 | 486.9 | 163.0 | 161.8 | 143.5 | 138.6 |
|  |  |  |  |  |  |  |  |  |
| A-doublet $_{Flk}$ | 539.3 | 530 | 494.3 | 485.9 | 162.8 | 161.4 | 142.8 | 138.0 |
| B-doublet $_{Flk}$ | 540 | 531.5 | 494.9 | 486.9 | 163.4 | 162.2 | 143.8 | 139.0 |

**Table S1.** Experimental binding energies for the mineral chips (Min) and the single flake (Flk) of franckeite.

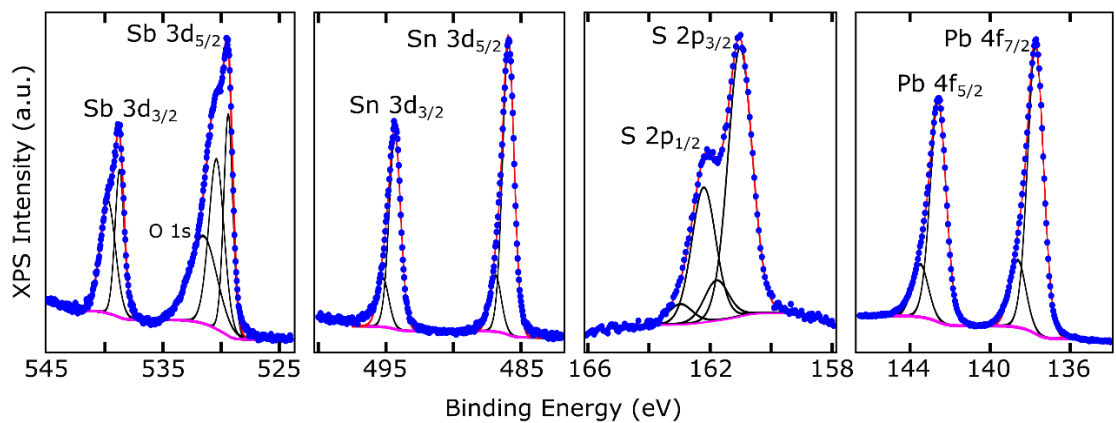

**Figure S6.** XPS spectra of Sb 3d, Sn 3d, S 2p and Pb 4f regions of a franckeite mineral chip.

As in the single flake, the best fit with the minor number of meaningful components results in two doublets for each of the measured element: we name "A-doublet" to the doublet with lower binding energy and "B-doublet" to the one with higher binding energy. The A-doublet of Pb4f (with Pb $4f_{7/2}$ at 137.7 eV) corresponds to the lead in the Q layer containing PbS,[6, 7] while the B-doublet (Pb $4f_{5/2}$ at 138.6 eV) can be attributed to a lead oxide component. In the Sn 3d core level, the A-doublet (Sn $3d_{5/2}$ at 485.9 eV) can be attributed to the state $Sn^{2+}$, and the B-doublet (Sn $3d_{5/2}$ at 486.9 eV) to $Sn^{4+}$: the difference in near 1 eV between A- and B-doublet is in accordance with the literature for these states.[6, 8, 9, 10]

But in the franckeite structural model[3] the $Sn^{2+}$ ion is a minority respect $Sn^{4+}$ and it is present in the Q layer while the $Sn^{4+}$ ion is majority and present in the H layer as $SnS_2$, whereas in our sample the $Sn^{2+}$ component is pretty much higher than $Sn^{4+}$. The fact that we see more amount of $Sn^{2+}$ than $Sn^{4+}$ can indicate that fracturing/exfoliation occurs preferentially leaving exposed the Q layer, explaining at the same time the appearance of the high binding energy component of Pb as surface atoms with a different chemical shift or a lead oxide component. The Sb as well can be fit too with two doublets, although there is a poor quality at the low binding energy side of Sb $3d_{5/2}$, that can indicate a surface state or defects produced by the fracturing. The B-doublet (Sb $3d_{5/2}$ at 530.4 eV) corresponds to $Sb^{3+}$ state in accordance with literature,[11] but the A-doublet (Sb $3d_{5/2}$ at 529.4 eV) that has lower binding energy and a high intensity must be attributed to Sb inclusions or substitutional defects. For S 2p, the A-doublet (S $2p_{3/2}$ at 161 eV) and the B-doublet (S $2p_{3/2}$ at 161.8 eV) are completely in accordance with the lead and tin sulfide.

From the area of the most intense peaks and the atomic sensitivity factor we can estimate the composition as $Pb_{25.6}Sn_{9.0}Fe_{0.7}Sb_{10.0}S_{53.7}Ag_{1.0}$. This is close to the composition found by Makovicky *et al.*[5] of $Pb_{21.7}Sn_{9.3}Fe_{4.0}Sb_{8.1}S_{56.9}$ but with some discrepancies, appearance of extra Sb in our samples, and the presence of Ag in higher amount than Fe.

**Liquid phase exfoliation**

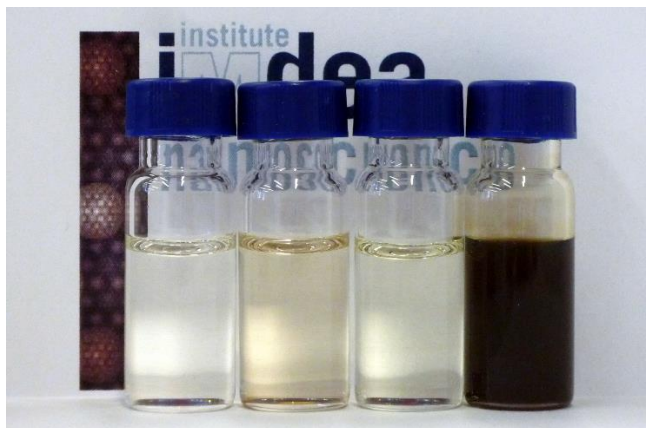

**Figure S7 | Franckeite colloidal dispersions.** From left to right: colloidal suspensions obtained after sonication of 0.1, 1, 10 and 100 mg·mL$^{-1}$ franckeite powder dispersions in NMP and centrifugation.

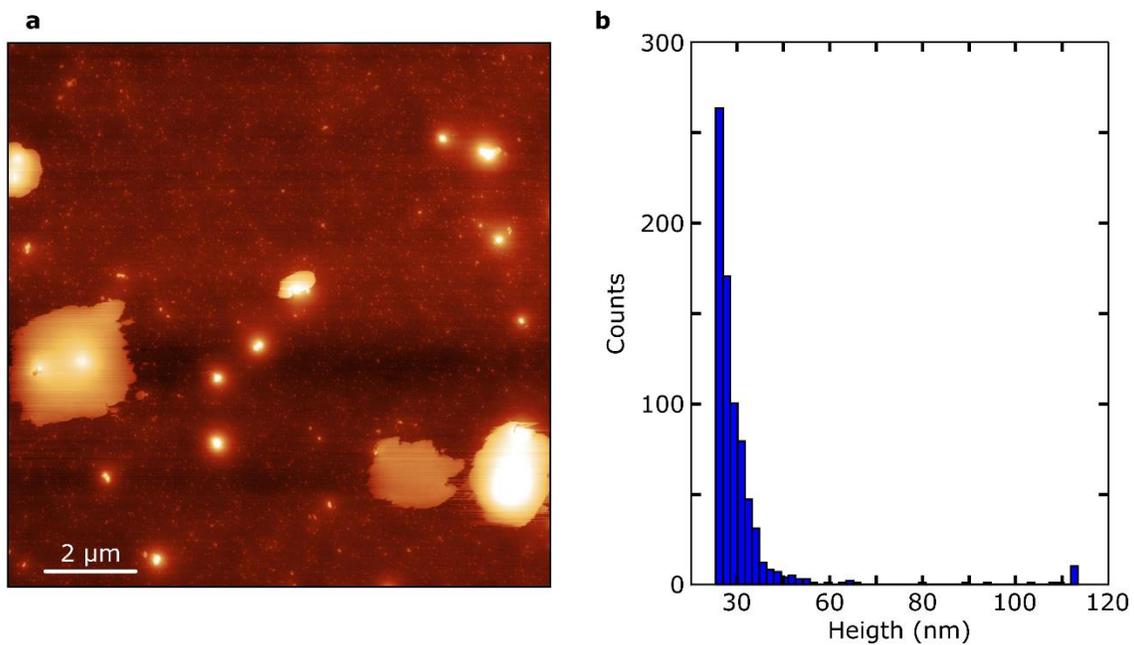

**Figure S8 | AFM of NMP-exfoliated franckeite colloidal dispersion prepared by sonication of a 100 mg·mL$^{-1}$ dispersion. a**, AFM micrograph of the sample dropcasted and dried over a freshly exfoliated mica substrate. **b**, Statistical analysis of the raw height data.

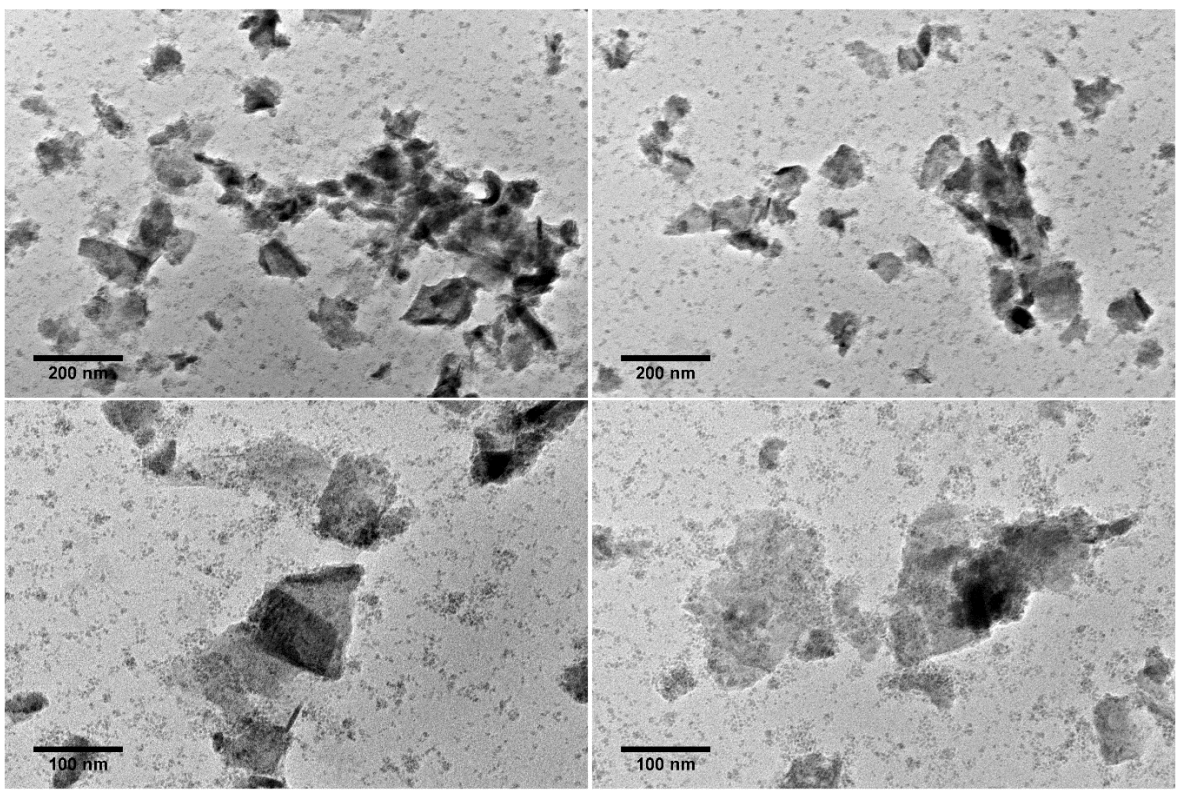

**Figure S9 | TEM images of franckeite nanosheets obtained in NMP from a 100 mg·mL$^{-1}$ powder dispersion.**

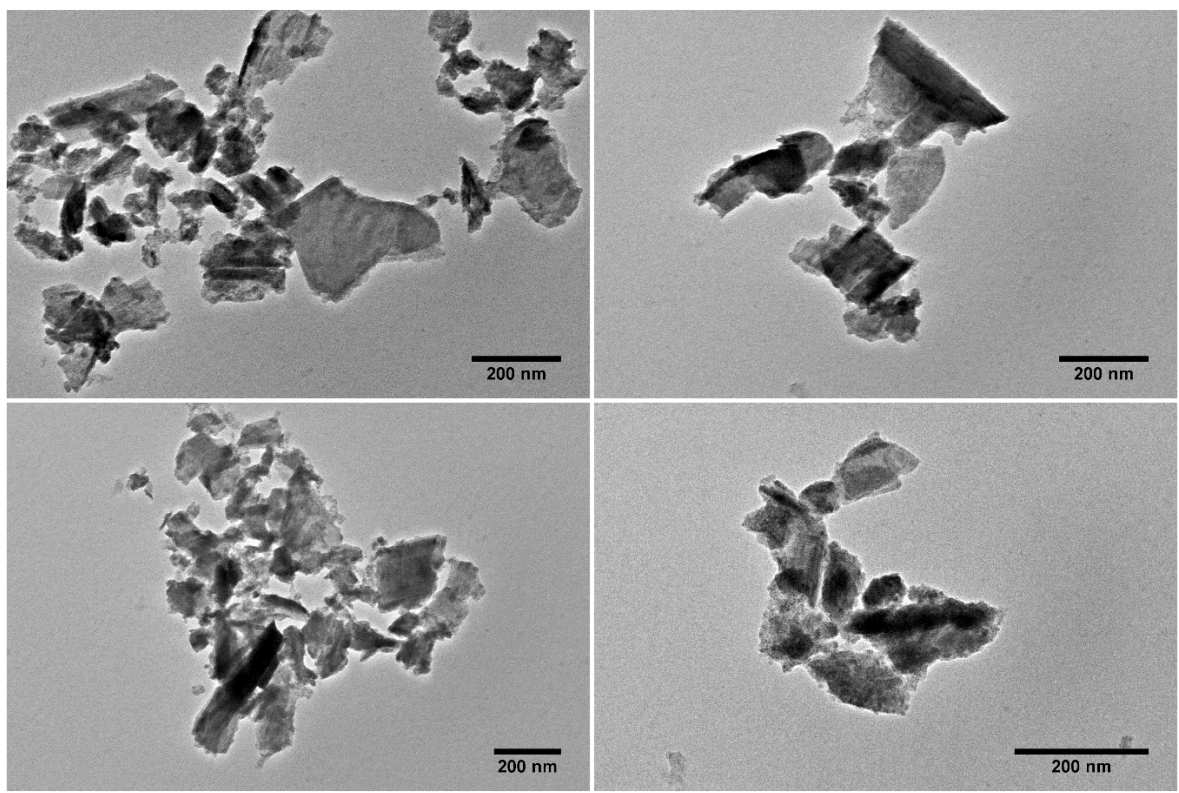

**Figure S10 | TEM images of franckeite nanosheets obtained in IPA/water (1/4, v/v) from a 1 mg·mL$^{-1}$ powder dispersion.**

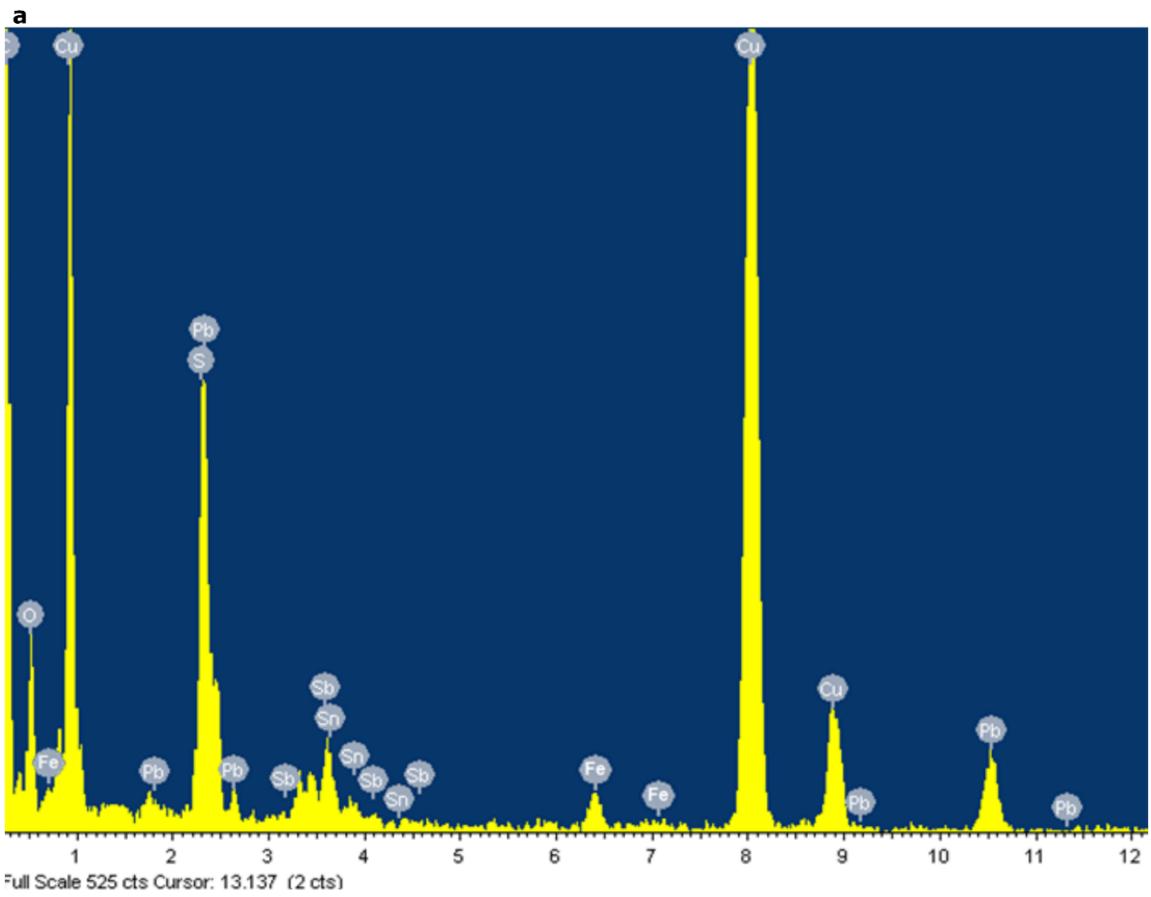
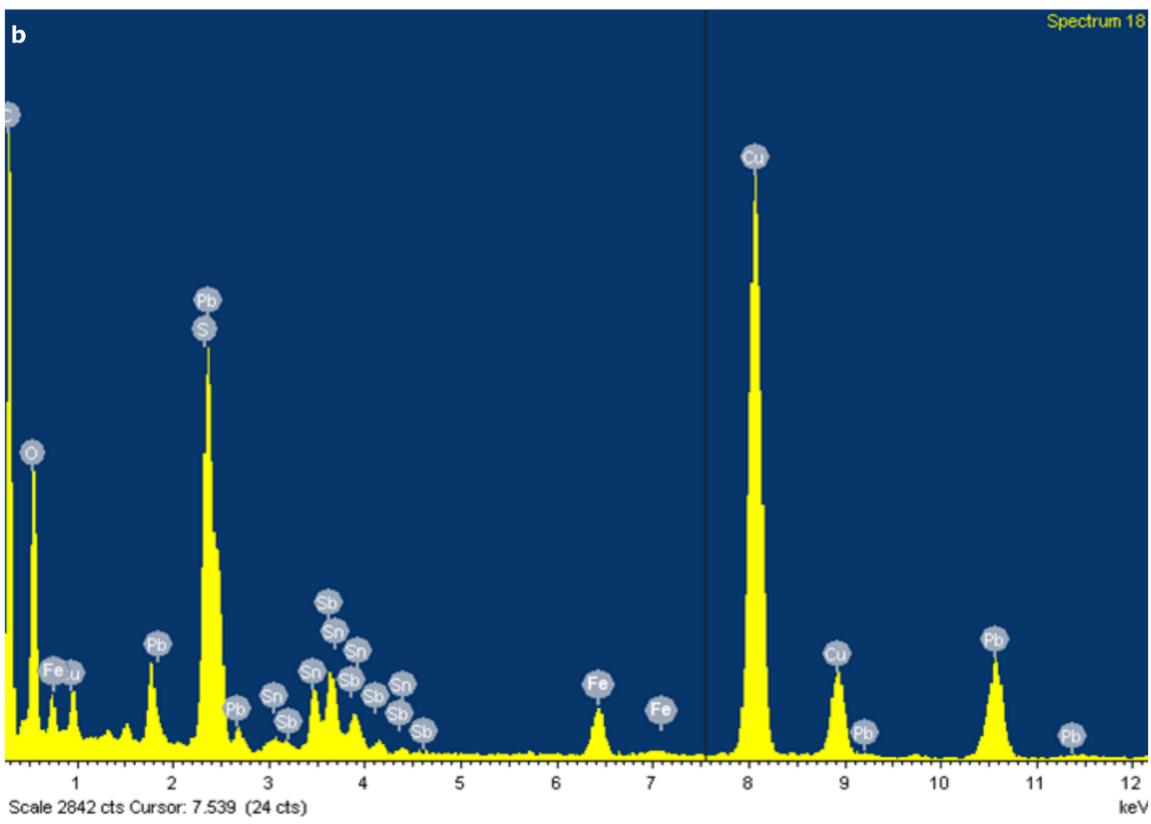

**Figure S11 | EDX spectra of exfoliated franckeite nanosheets. a**, obtained in NMP from a 100 mg·mL$^{-1}$ powder dispersion; **b**, obtained in IPA/water 1/4 from a 1 mg·mL$^{-1}$ powder dispersion.

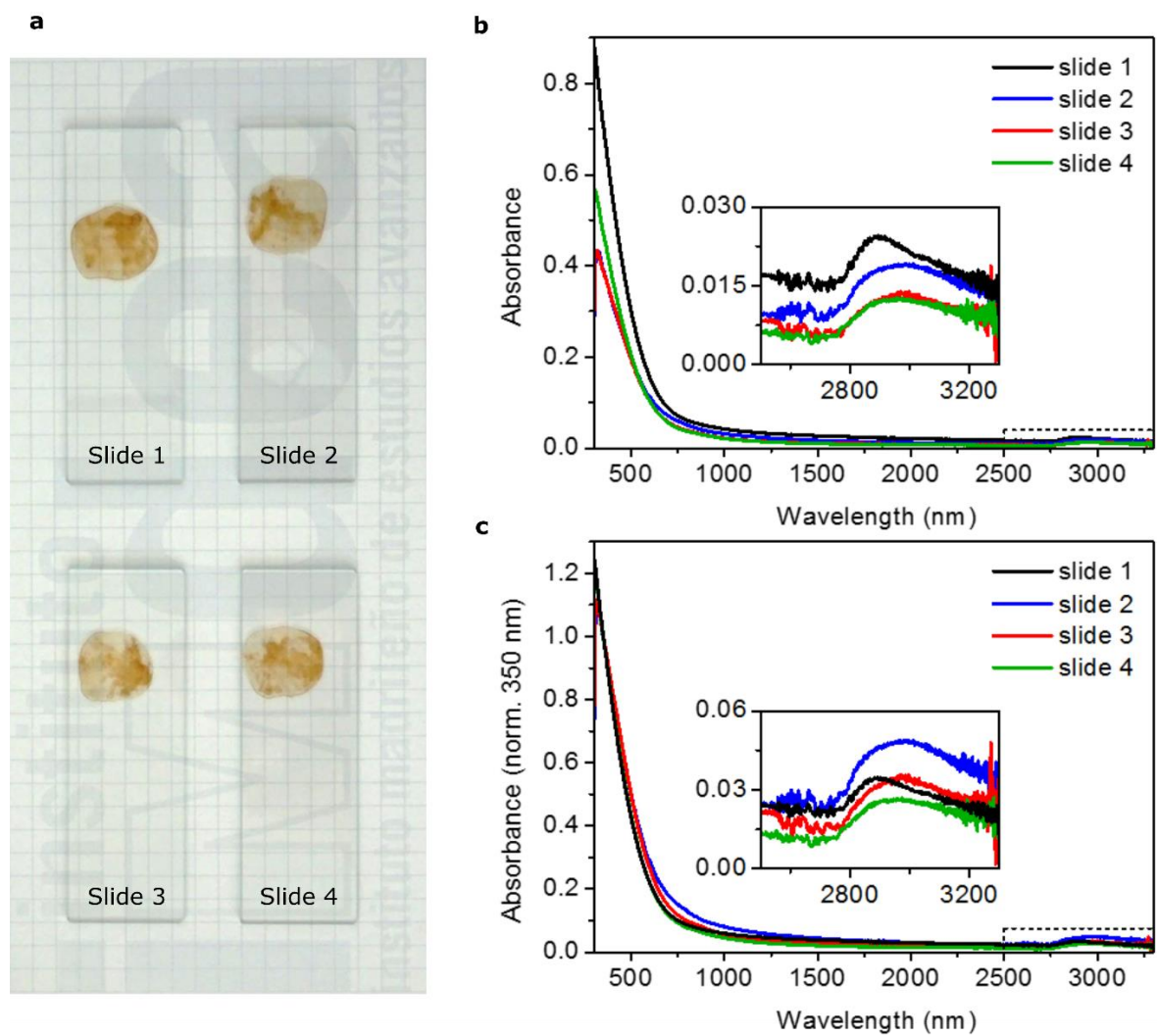

**Figure S12 | UV-Vis-NIR spectra of exfoliated franckeite (obtained from the 100 mg.mL$^{-1}$ powder dispersion) drop-casted and dried onto glass slides. a**, Exfoliated sample films deposited onto the four different glass slides used for the measurements. **b**, Absorption spectra of the sample shown in the main text (slide 1, black line), and of the three replicates performed with other three different glass slides (slides 2, 3 and 4; blue, red

and green lines respectively); inset: zoom in the 2500 nm-3300 nm region. **c**, Absorption spectra normalized at 350 nm for the four slides presented in **c**; inset: zoom in the 2500 nm-3300 nm region.

**Powder and liquid-phase-exfoliated franckeite Raman spectra interpretation**

The band at 66 cm$^{-1}$ corresponds to an acoustic phonon mode in PbS,[12,13,14] whereas the band at 318 cm$^{-1}$ is a clear signature of the $A_{1g}$ mode in SnS$_2$.[15,16,17,18] Other bands at 145, 194 cm$^{-1}$, and the 400-650 cm$^{-1}$ shoulder are most likely due to a superposition of respective phonon modes of both PbS (combinations and overtones of acoustic and optical modes) [12,13,14,19] and SnS$_2$ (including the $E_g$ mode and second-order effects). [15,16,17,18] As proposed in other works on misfit compounds,[14] the last band at 253 cm$^{-1}$ could result from the combination of phonon modes of both layer types. Slight shifts are to be noted regarding the parent structures; they can be attributed to deviations in terms of structure[4,20,21] and composition (*e.g.* Sb and Fe identified by TEM/EDX)[22] between the idealized model and natural franckeite. In spite of its weakness, the interlayer van der Waals interaction can also cause some noticeable changes in the respective vibration properties of each lattice.[14]

**Table S2 | Interpretation of Raman spectra of franckeite (powder and exfoliated particles).**

| Raman shift (cm$^{-1}$) | Phonon mode attribution | Homogeneous material reference | | | |
|---|---|---|---|---|---|
| | | Raman shift (cm$^{-1}$) | Compound | State | Literature citation |
| 66 | SA | 68 | PbS | nanocrystal | 12, 13 |
| | 2TA | 73 | | (PbS)$_{1.18}$(TiS$_2$)$_2$ misfit | 14 |
| 145 | 2$^{nd}$ order effect | 131 | SnS$_2$ | nanosheets | 15 |
| | | 135 | | nanocrystallites | 16 |
| | | 140.5 | | bulk | 17 |
| | TA + TO | 151 | PbS | (PbS)$_{1.18}$(TiS$_2$)$_2$ misfit | 14 |
| | | 154 | | bulk single crystal | 19 |
| 194 | LO(Γ) | 203 | PbS | (PbS)$_{1.18}$(TiS$_2$)$_2$ misfit | 14 |
| | | 204 | | bulk single crystal | 19 |
| | | 215 | | nanocrystal | 12, 13 |
| | E$_g$ | 198 | SnS$_2$ | nanocrystallites | 16 |
| | | 205 | | bulk | 17 |
| | | 212 | | nanosheets | 15 |
| | | 215 | | film (CVD) | 18 |
| 253 | combination | | PbS + SnS$_2$ | | 14 |
| 318 | A$_{1g}$ | 309 | SnS$_2$ | nanocrystallites | 16 |
| | | | | nanosheets | 15 |
| | | 312 | | film (CVD) | 18 |
| | | 315 | | bulk | 17 |
| 400-650 | 2LO(Γ) | 412 | PbS | (PbS)$_{1.18}$(TiS$_2$)$_2$ misfit | 14 |
| | | 415 | | nanocrystal | 13 |
| | | 454 | | bulk single crystal | 19 |
| | 3LO(Γ) | 630 | | nanocrystal | 13 |
| | | 632 | | (PbS)$_{1.18}$(TiS$_2$)$_2$ misfit | 14 |
| | 2$^{nd}$ order effect | 450-650 | SnS$_2$ | nanocrystallites | 16 |
| | | | | nanosheets | 15 |

Abbreviations: phonon modes TO: transverse optical; TA: transverse acoustic; LO: longitudinal optical; SA: spheroidal acoustic. (Γ): Γ point of the Brillouin zone.

**Thermopower**

We have studied the thermopower of bulk franckeite by using a modified scanning tunneling microscope (STM) which allows to locally measure the thermpower of a surface. The gold STM tip, which is mechanically cut, is moved towards the sample for 70-90 nm to form a large contact. The sample is kept at room temperature while the tip is resistively heated up to a temperature gradient of ~35K. In addition to the applied bias voltage, the temperature gradient gives rise to a voltage offset $-S\Delta T$, where S is the thermopower of the franckeite (Fig. S13).

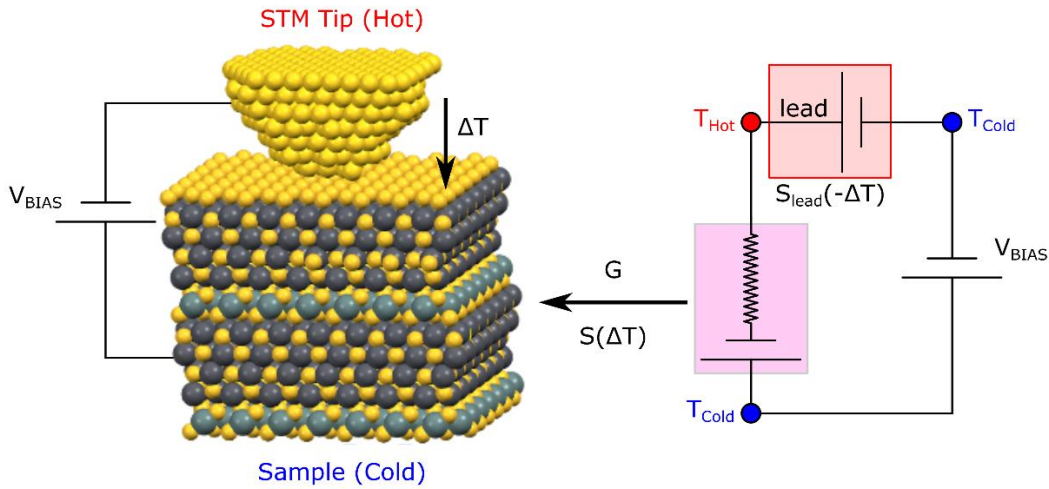

**Figure S13 | Scanning tunneling microscope for measuring thermopower.** Equivalent thermal circuit of the setup for measuring the thermopower. The sample is kept at ambient temperature $T_{Cold}$ while the tip is heated to a temperature of $T_{Hot} = T_{Cold} + \Delta T$. S is the thermopower of franckeite and $S_{lead}$ is the thermopower of the tip connecting lead. $V_{BIAS}$ is the bias voltage applied at the sample.

This offset can be measured from a current-voltage curve by finding the voltage that makes the current zero. The slope of the current-voltage curve gives the conductance (Fig. S14). [23, 24]

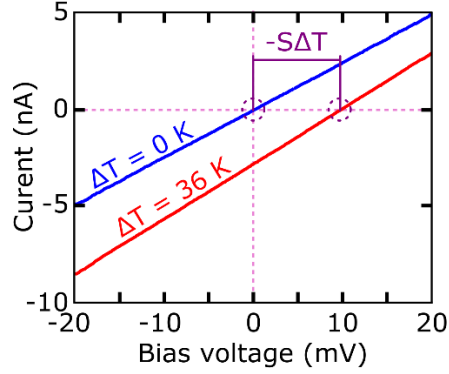

**Figure S14.** Current-voltage curves acquired on franckeite at $\Delta T = 0$ K (blue) and $\Delta T = 36$ K (red) showing the thermal voltage $-S\Delta T$ due to the temperature gradient in the sample.

By heating the tip we not only establish a temperature difference between the tip and the substrate but also a temperature gradient across the tip-connecting lead, which gives rise to an additional thermoelectric voltage. The equivalent circuit for measuring the thermopower is shown in Fig. S13. Considering the equivalent circuit, we can write $I = G\,(V_{bias}+S_{lead}\,\Delta T - S\Delta T)$, where $S$ and $S_{lead}$ is the thermopower of the franckeite and the tip connecting lead respectively, and $\Delta T = T_{Hot}-T_{Cold}$ is the temperature gradient, with $T_{Hot}$ and $T_{Cold}$ being the temperatures of the tip and the franckeite, respectively. It has been shown previously that $S_{lead} = 0.05\ \mu V \cdot K^{-1}$, [2] which is negligible in comparison with the thermopower of franckeite. In addition, the resulting value of the measured thermopower is the difference between the thermopower of gold and franckeite, but since the thermopower of gold is much smaller than the franckeite, its contribution is negligible.

To obtain the average thermopower of franckeite we collected data of 85 traces in a histogram and we found a positive value of $S = 264$ µV·K$^{-1}$, indicating that franckeite is a p-type material. In addition, the mean conductance was found to be $G = 0.0032\ G_0$ ($R=4$ MΩ), where is the $G_0$ quantum of conductance. The thermal conductivity of gold is expected much larger than that of franckeite when making a large contact, thus most of the temperature drops in the franckeite itself and consequently the thermopower measured is the bulk thermopower of the franckeite.[25]

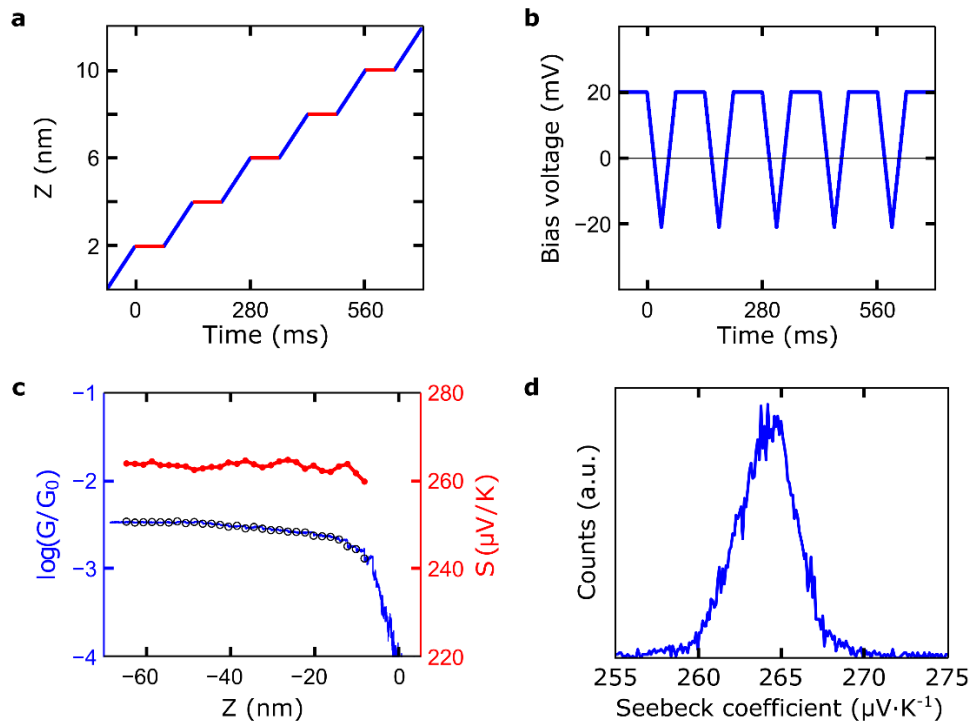

**Figure S15 | Measurement of the thermopower. a** and **b,** Tip displacement $z$ and applied bias voltage $V_{BIAS}$, respectively, as a function of time. The bias voltage is maintained at a fixed value 20 mV during the tip motion and every 2 nm it is swept between ±20 mV while

the tip is stationary (examples of *IV* curves are shown in Fig. S14). **c,** Characteristic example of conductance (blue) and thermopower (red) acquired simultaneously for approach of the Au tip and large contact formation with the franckeite. The temperature gradient across the junction was set to $\Delta T \approx 36$ K. Note that the thermopower and conductance stabilizes when a large contact is established, indicating that indeed the thermopower value measured is the bulk value. Black open circles indicate the points where the tip motion stops and the current-voltage curves were recorded. **d,** Histogram of thermopower of franckeite, out of 85 traces at $\Delta T \approx 36$ K.

**Optical microscopy characterization of a franckeite photodetector**

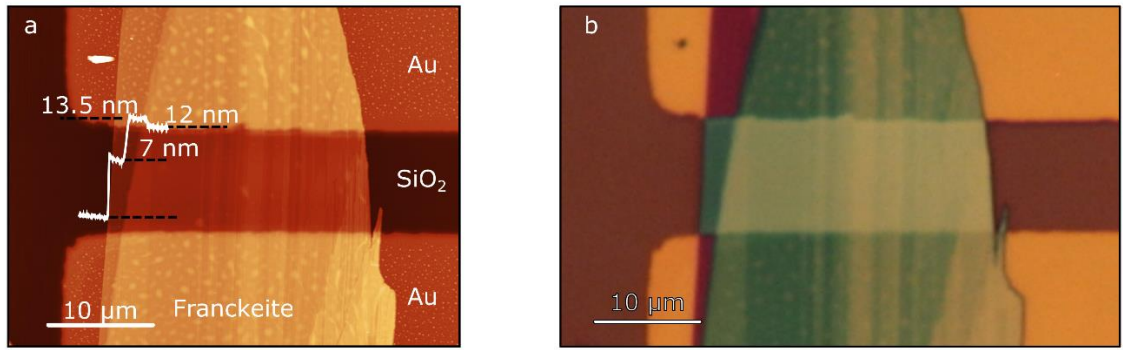

**Figure S16. a** and **b,** AFM and optical microscopy image of the franckeite device characterized in Figure 4 of the main text, respectively.

## MoS$_2$-franckeite p-n junction

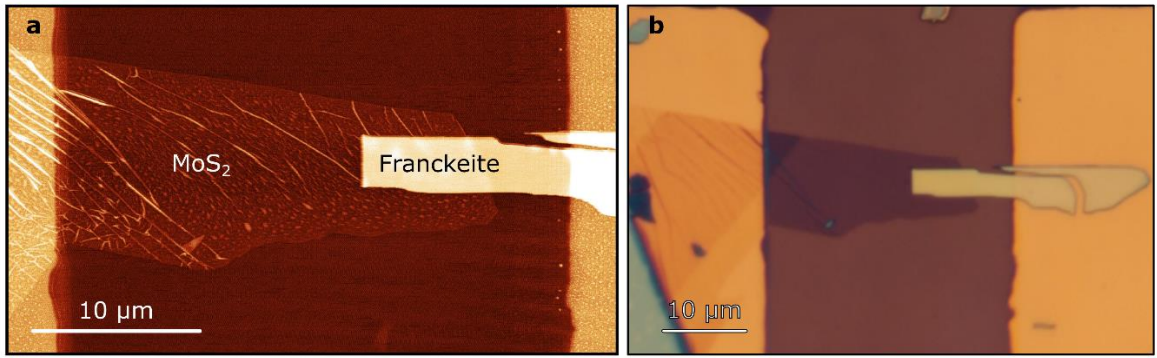

**Figure S17. a** and **b,** AFM and optical microscopy image of the p-n junction (MoS$_2$ - franckeite) characterized in Figure 5 of the main text.

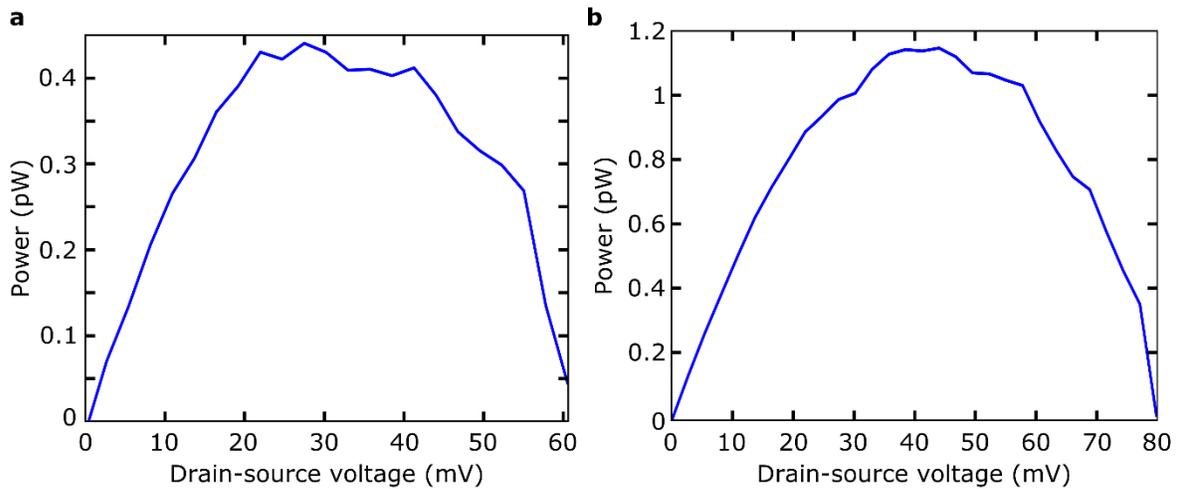

**Figure S18** p-n junction electrical power harvested in the device, calculated as $P_{el} = |I_{ds}| \cdot V_{ds}$ upon illumination with a laser spot of **a** 940 nm wavelength with a $P_{el,max} \sim 0.5$ pW and **b** 885 nm wavelength with a $P_{el,max} \sim 1.2$ pW.

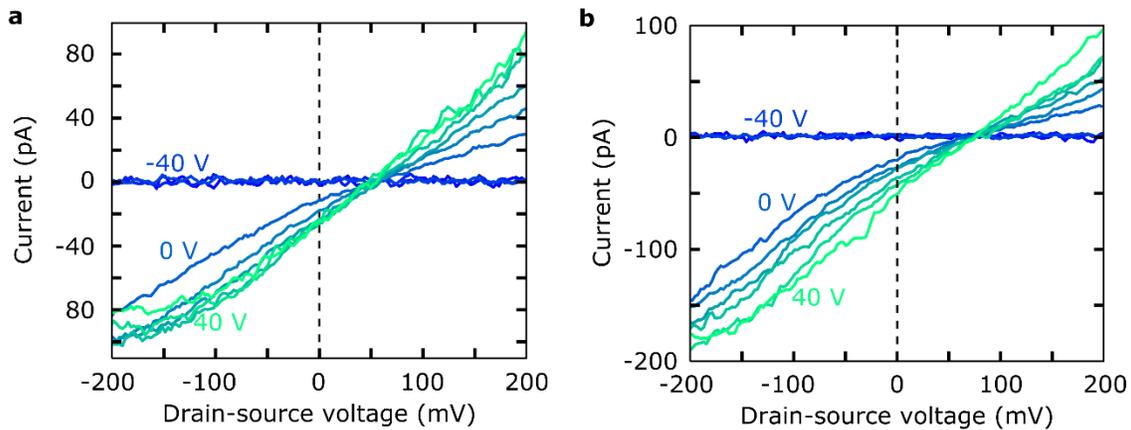

**Figure S19.** p-n junction current-voltage characteristics for different back-voltages ranging from -40 V to 40 V upon illumination with **(a)** 940 nm wavelength and **(b)** 885 nm wavelength.

## Supporting Information references

1. Crampin S, Jensen H, Kröger J, Limot L, Berndt R. Resonator design for use in scanning tunneling spectroscopy studies of surface electron lifetimes. *Physical Review B* 2005, **72**(3): 035443.

2. Maccariello D, Garnica M, Niño MA, Navío C, Perna P, Barja S, *et al.* Spatially Resolved, Site-Dependent Charge Transfer and Induced Magnetic Moment in TCNQ Adsorbed on Graphene. *Chem Mater* 2014, **26**(9): 2883-2890.

3. Wang S, Kuo K. Crystal lattices and crystal chemistry of cylindrite and franckeite. *Acta Crystallogr Sect A: Found Crystallogr* 1991, **47**(4): 381-392.

4. Williams TB, Hyde BG. Electron microscopy of cylindrite and franckeite. *Physics and Chemistry of Minerals* 1988, **15**(6): 521-544.

5. Makovicky E, Petříček V, Dušek M, Topa D. The crystal structure of franckeite, Pb21. 7Sn9. 3Fe4. 0Sb8. 1S56. 9. *Am Mineral* 2011, **96**(11-12): 1686-1702.

6. Shalvoy RB, Fisher GB, Stiles PJ. Bond ionicity and structural stability of some average-valence-five materials studied by x-ray photoemission. *Phys Rev B* 1977, **15**(4): 1680-1697.

7. Zhang YC, Qiao T, Hu XY, Wang GY, Wu X. Shape-controlled synthesis of PbS microcrystallites by mild solvothermal decomposition of a single-source molecular precursor. *J Cryst Growth* 2005, **277**(1): 518-523.

8. Cruz M, Morales J, Espinos JP, Sanz J. XRD, XPS and 119Sn NMR study of tin sulfides obtained by using chemical vapor transport methods. *J Solid State Chem* 2003, **175**(2): 359-365.

9. Zhang YC, Li J, Xu HY. One-step in situ solvothermal synthesis of SnS2/TiO2 nanocomposites with high performance in visible light-driven photocatalytic reduction of aqueous Cr(VI). *Applied Catalysis B: Environmental* 2012, **123–124:** 18-26.

10. Cheng S, Conibeer G. Physical properties of very thin SnS films deposited by thermal evaporation. *Thin Solid Films* 2011, **520**(2): 837-841.

11. Zakaznova-Herzog VP, Harmer SL, Nesbitt HW, Bancroft GM, Flemming R, Pratt AR. High resolution XPS study of the large-band-gap semiconductor stibnite (Sb2S3): Structural contributions and surface reconstruction. *Surf Sci* 2006, **600**(2): 348-356.

12. Krauss TD, Wise FW. Coherent Acoustic Phonons in a Semiconductor Quantum Dot. *Physical Review Letters* 1997, **79**(25): 5102-5105.

13. Krauss TD, Wise FW. Raman-scattering study of exciton-phonon coupling in PbS nanocrystals. *Physical Review B* 1997, **55**(15): 9860-9865.

14. Ovsyannikov SV, Shchennikov VV, Cantarero A, Cros A, Titov AN. Raman spectra of $(PbS)_{1.18}(TiS_2)_2$ misfit compound. *Mater Sci Eng A* 2007, **462**(1–2): 422-426.

15. Kang J-G, Lee G-H, Park K-S, Kim S-O, Lee S, Kim D-W, *et al.* Three-dimensional hierarchical self-supported multi-walled carbon nanotubes/tin(IV) disulfide nanosheets heterostructure electrodes for high power Li ion batteries. *Journal of Materials Chemistry* 2012, **22**(18): 9330-9337.

16. Wang C, Tang K, Yang Q, Qian Y. Raman scattering, far infrared spectrum and photoluminescence of SnS2 nanocrystallites. *Chemical Physics Letters* 2002, **357**(5–6): 371-375.

17. Smith AJ, Meek PE, Liang WY. Raman scattering studies of $SnS_2$ and $SnSe_2$. *J Phys C* 1977, **10**(8): 1321-1333.

18. Price LS, Parkin IP, Hardy AME, Clark RJH, Hibbert TG, Molloy KC. Atmospheric Pressure Chemical Vapor Deposition of Tin Sulfides (SnS, $Sn_2S_3$, and $SnS_2$) on Glass. *Chemistry of Materials* 1999, **11**(7): 1792-1799.

19. Smith GD, Firth S, Clark RJH, Cardona M. First- and second-order Raman spectra of galena (PbS). *Journal of Applied Physics* 2002, **92**(8): 4375-4380.

20. Wang S, Kuo KH. Crystal lattices and crystal chemistry of cylindrite and franckeite. *Acta Crystallographica Section A* 1991, **A47**(4): 381-392.

21. Makovicky E, Petříček V, Dušek M, Topa D. The crystal structure of franckeite, $Pb_{21.7}Sn_{9.3}Fe_{4.0}Sb_{8.1}S_{56.9}$. *American Mineralogist* 2011, **96**(11-12): 1686-1702.

22. Mehner H. Mössbauer Investigations on Minerals of the Franckeite - Cylindrite Group. *Hyperfine Interactions* 1998, **112**(1): 239-241.

23. Evangeli C, Gillemot K, Leary E, González MT, Rubio-Bollinger G, Lambert CJ, *et al.* Engineering the Thermopower of C60 Molecular Junctions. *Nano Letters* 2013, **13**(5): 2141-2145.

24. Rincon-Garcia L, Ismael AK, Evangeli C, Grace I, Rubio-Bollinger G, Porfyrakis K*, et al.* Molecular design and control of fullerene-based bi-thermoelectric materials. *Nat Mater* 2016, **15**(3)**:** 289-293.

25. Evangeli C, Matt M, Rincón-García L, Pauly F, Nielaba P, Rubio-Bollinger G*, et al.* Quantum Thermopower of Metallic Atomic-Size Contacts at Room Temperature. *Nano Letters* 2015, **15**(2)**:** 1006-1011.